%% file: jinst_dipcoating.tex
\documentclass[hyper]{JINST}
\pdfoutput=1

\usepackage{float}
\usepackage{graphicx}
\usepackage{bm}
\usepackage{subfig}
\usepackage{xspace}
\usepackage{lineno}
\usepackage{multirow}
\usepackage{eqnarray,amsmath}

\title{Improved TPB-coated Light Guides for Liquid Argon TPC Light Detection Systems}

\author{Z. Moss, L. Bugel, G. Collin, J.M. Conrad, B. J. P. Jones, J. Moon, M. Toups, T. Wongjirad\\
Physics Dept., Massachusetts Institute of Technology, Cambridge, MA 02139\\}

\abstract{
  Scintillation light produced in liquid argon (LAr) must be shifted from 
  128 nm to visible wavelengths in light detection systems used 
  for liquid argon time-projection chambers (LArTPCs).  To date, LArTPC light collection systems have employed tetraphenyl butadiene
  (TPB) coatings on photomultiplier tubes (PMTs) or plates placed in
  front of the PMTs.   Recently, a new approach using TPB-coated light
  guides was proposed.    In this paper, we 
  report on light guides with improved attenuation lengths above 100
  cm when measured in air.  This is an important step in the development of meter-scale light guides for future LArTPCs.  Improvements come from using a new acrylic-based coating, diamond-polished cast UV transmitting acrylic bars, and a hand-dipping technique to coat the
  bars. We discuss a model for connecting bar response in air to response in liquid argon and compare this to data taken in liquid argon.  The good agreement between the prediction of the model and the measured response in liquid argon demonstrates that characterization in air is sufficient for quality control of bar production.    This model can be used in simulations of light guides for future experiments.
}


\begin{document}


\section{Introduction}
\label{sec:intro}

Liquid Argon Time Projection Chambers (LArTPCs) are a class of detectors that hold great promise for the future of neutrino physics. 
One advantage these detectors have is their ability to produce high resolution images of charged particle tracks traveling through them.  
With such images, future neutrinos experiments aim to better distinguish neutrino events from background.
Current neutrino LArTPCs consist of time projection chambers (TPCs) that are several squaremeters in cross section and many meters in length. Each is housed in a cryostat filled with ultra high purity liquid argon (LAr) that ionizes when neutrino (and background) 
events produce charged particles that traverse it.  An applied electric field drifts the ionization electrons to an anode plane over the course of a few milliseconds, where their charge and arrival time are read out by sense wires. 
In addition to ionization electrons, charged particles traveling through the LAr produce vacuum ultraviolet (VUV) scintillation light, which has a nanosecond-scale fast component (6 ns) and a microsecond-scale slow component (1.6 $\mu$s).  This scintillation light can be used in a number of ways: to trigger detector readout, determine the absolute drift time of non-accelerator events, to reject cosmic backgrounds, and to complement charge-based event reconstruction.  Therefore, a major effort in the research and development (R\&D) of such light collection systems is currently underway \cite{Bugel:2011fw, Baptista2013, Buchanon2013, Szelc:2013ht}.   One of the primary challenges in this effort is the detection of the LAr scintillation light itself, which peaks around 128 nm.  This
wavelength cannot be observed directly by commonly used photodetectors such as PMTs.  Therefore, one direction in the R\&D for LArTPC light collection
systems is to develop efficient, inexpensive systems that respond to this VUV light. 
 

One cost-effective approach is to use a
flat-profile system based on light-guiding bars that are assembled
into panels.   These panels can be inserted into dead regions, such as the volume behind
a TPC wire chamber located along the wall of the LArTPC cryostat, or
in the dead region between two-sided TPC wire planes interspersed
throughout the liquid.  The former represents the proposed LAr1ND design~\cite{Adams:2013uaa}, and the latter represents the proposed design for
the ELBNF detector~\cite{LBNEdesign} .  
The light guide bars operate by first shifting the wavelength of scintillaton photons from the liquid argon to a longer wavelength. 
This shift is done by a coating of TPB that, when excited, isotropically emits photons over a spectrum peaked around 425 nm\cite{Gehman:2011xm}.
A portion of the reemitted light is then guided by total internal reflection to the ends of the bar. 
Here, one or more photon detectors, such as PMTs or silicon photomultipliers (SiPMs), detect the wavelength-shifted light signal.

An initial proposal for a light guide-based system was presented in
Ref.~\cite{Bugel:2011fw}.  Since then, many improvements have
been made in light guide technology. Here we report on production techniques capable of achieving an attenuation length of 1 m in air, which is a substantial improvement over the original design.
Improvements are attributed to the use of cast and diamond-polished bars in lieu of extruded bars; 
UV transmitting (UVT) acrylic rather than UV protected acrylic;  an
improved coating formula (provided below); and a dip-coating process in place of a hand-painting method for coating the bars. Many steps in the process of
improving the light guides are reported in Ref.~\cite{Baptista:2012bf}. Here, we describe the outcome of this R\&D campaign without discussing the intermediate steps. 

In what follows, we describe the construction of the acrylic bars and the process with which we coat them in Sec.~\ref{sec:bars} and~\ref{sec:coating}.  Measurements quantifying the behavior of the bars in both air and liquid argon are described in Sec.~\ref{sec:attair} and~\ref{sec:attlar}, respectively.  We then present a model connecting the measured behavior of the bars in air and liquid argon in Sec.~\ref{sec:model} and give our conclusions in Sec.~\ref{sec:conclusions}.

\section{Construction of the Light Guide Bars}
\label{sec:bars}

Cast UVT acrylic sheets (Lucite UTRAN, Plaskolite Inc.) were
laser-cut into bars measuring $20'' \times 1'' \times 1/4''$ and then
diamond-polished on the sides and ends by Altec Plastics.   We have
observed visible crazing on the sides and ends
of the polished bars covered with our wavelength shifting coating after immersion in LAr. 
This crazing is expected to lower the overall light output and shorten the
attenuation length by introducing defects from which light in the bar will
scatter. In order to reduce the amount of crazing, the bars are annealed
after cutting/polishing and before coating with the wavelength shifter. 

The bars are annealed using a Steinel HL 2010 E electronic heat gun
which warms bar in an insulated tube for 3 hours at 230 degrees Fahrenheit.
The temperature is then stepped down in 10-degree increments every 10 minutes to 120
degrees. The bars are then removed and allowed 30 minutes to cool to room
temperature. After this process is complete, the bars are wiped down
on all surfaces with pure (azeotropic) ethanol and handled exclusively with nitrile gloves thereafter. 

\section{Coating with Wavelength Shifter}
\label{sec:coating}

The wavelength shifting coating applied to the surface of the bars has evolved
through many cycles of experimentation. The current formula contains four compounds: 
toluene, as a solvent; TPB, as a wavelength shifter; acrylic, as a binder; and ethanol, 
as a surfactant. 
Dissolving acrylic into the liquid solution allows the formation of a thin film consisting of a durable, clear layer of TPB-rich plastic deposited on the surface of the bar. The addition of ethanol helps us achieve a smooth coat with minimal running or beading.

To produce one $60~$mL batch of coating solution, $0.5~$g of scintillation grade TPB (Sigma-Aldrich) is combined with $1.0~$g of UVT acrylic pellets (Plaskolite) in $50$~mL of ACS spectrophotometric grade toluene, and spun in an orbital shaker to dissolve overnight. Once the TPB and UVT acrylic are fully dissolved, $10$~mL of pure ethanol is added, and the solution is again mixed for $10$ minutes.

\begin{figure}[t] 
\centering 
\includegraphics[scale=0.45]{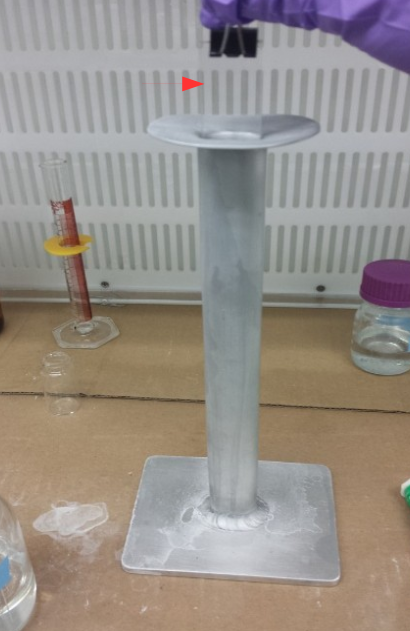} 
\caption{A bar held by a clip in a shortened prototype (with similar geometry) of the dipping vessel. The red arrow points to the transparent bar.}
\label{fig:short_dipper}
\end{figure}

The coating solution is then poured into the dip coating vessel, shown in Fig.~\ref{fig:short_dipper}. This vessel is a roughly elliptical tube measuring 51.4 cm in length, 
attached at the bottom to a stable base, and at the top to a wide lip to ease fluid transfer. 

A clean bar is then inserted into the vessel, ideally displacing the coating solution to a height not more than a few centimeters short of the end of the bar. The solution should not reach the dry end of the bar, as this end is polished, and any chemical attack could scatter light that would otherwise exit the end of the bar into a photodetector.  In the case of the bars presented in section~\ref{sec:ezb}, the soak time is 5 minutes. The bars are then quickly withdrawn from the solution and hung vertically by a clip from the dry end.  They are left to air dry for a minimum of 30 minutes under a fume hood.

Compared to a previous hand-painting method~\cite{Bugel:2011fw}, this dip-coating technique visibly improved the clarity and uniformity of the coating, as well as the scalability of the bar coating process.  

\section{Attenuation Length Measurements in Air \label{sec:attair}}

We measure the attenuation length of the coated acrylic bars in air at
room temperature using a pulsed 286 nm (11.6 nm FWHM) UV LED.
The LED is positioned at points along the bar by a
computer-controlled stepper motor to an accuracy of $\pm 3~$mm.  
This uncertainty was measured by repeatedly deploying the LED to 
a given position and measuring the distance with a tape measure.

\begin{figure}[t] 
\centering 
\includegraphics[scale=0.45]{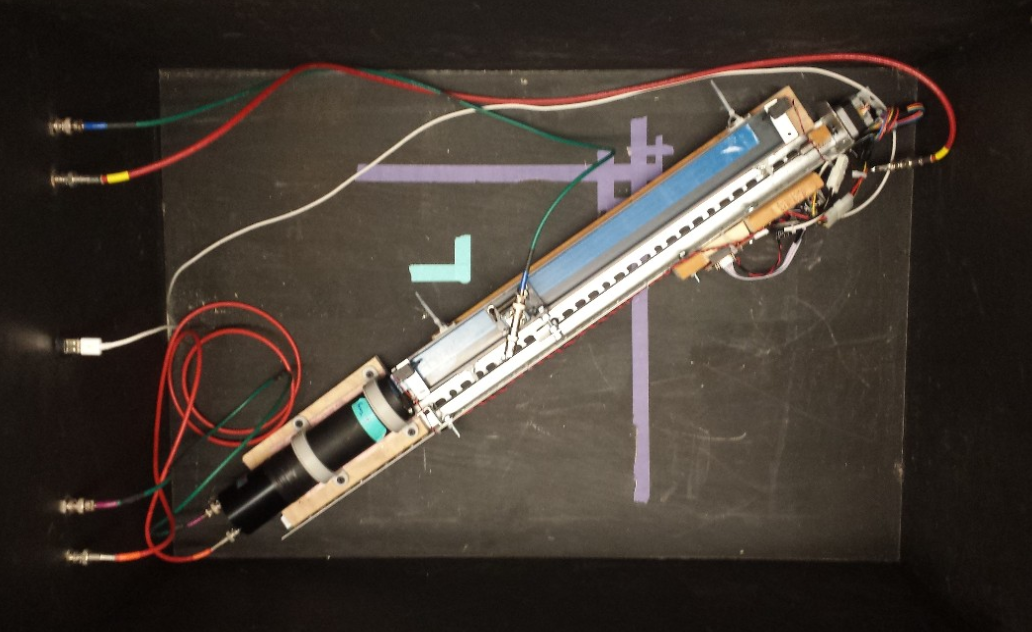} 
\caption{The attenuation length measurement apparatus (see text for details).}
\label{fig:led_setup}
\end{figure}

A 2'' Hamamatsu R1828-01 PMT biased at -1980 V is positioned at one end of the bar and detects any emerging light.  The entire apparatus was
housed in a dark box with black walls to mitigate the effects of stray light.  Figure~\ref{fig:led_setup} shows a picture of the setup.

Each LED run starts at the point nearest to the PMT and moves towards the
far end of the bar.  The PMT and LED are turned on 5 minutes prior to the 
acquisition and remain on for the duration of the run.  At each point along 
the bar, roughly 60,000 PMT waveforms are recorded
in coincidence with the LED pulse.  After making one pass towards the far 
end of the bar, the LED moves back along the bar towards the PMT,
repeating each measurement.  The LED is pulsed at a constant voltage,
pulse width (120 ns), and rate (1 kHz) from bar to bar, so these data can be used to extract the relative
brightnesses as well as the optical attenuation lengths of each
bar. In this way, we are able to benchmark the bars and examine the
effects of variations in coating processes on bar quality.

Readout of the PMT is performed using an 8-bit Alazar Tech ATS9870
digitizer. The ADC range is set to a total voltage scale of $\pm$1 V, and a trigger
is produced by a negative pulse with an amplitude that exceeds $130$~mV. When a trigger is produced, 128
pre-trigger samples and 384 post-trigger samples are recorded at a
sampling rate of 1 GS/s, leading to a total waveform spanning 512 ns.

In total, three sets of annealed, dip-coated, UVT bars were produced using two batches of coating solution.  For the first two sets, the coating solution came from a single batch of wavelength shifter divided into two volumes. The set dipped in the first volume is referred to here as ``Set 1'' and the set dipped in the second volume as ``Set 2''.  Five bars were produced for both Set 1 and Set 2, and in each set the bars were dipped successively in the same volume of solution without replenishment of the fluid.  This was done in order to investigate the effect of repeated dipping on the total brightness and attenuation length of the bar.   The bars in the third set were made with a different batch of coating solution but with the same formula. The third set was reserved for a study of the attenuation in air for different orientations of the bar (Section~\ref{subsec:orientation_air}) and to compare the light output of each bar in both air and liquid argon (Section~\ref{sec:attlar}).

\subsection{Attenuation Length and Extrapolated Zero Brightnesses \label{sec:ezb}}

\begin{figure}[t] 
\centering 
\includegraphics[scale=0.45]{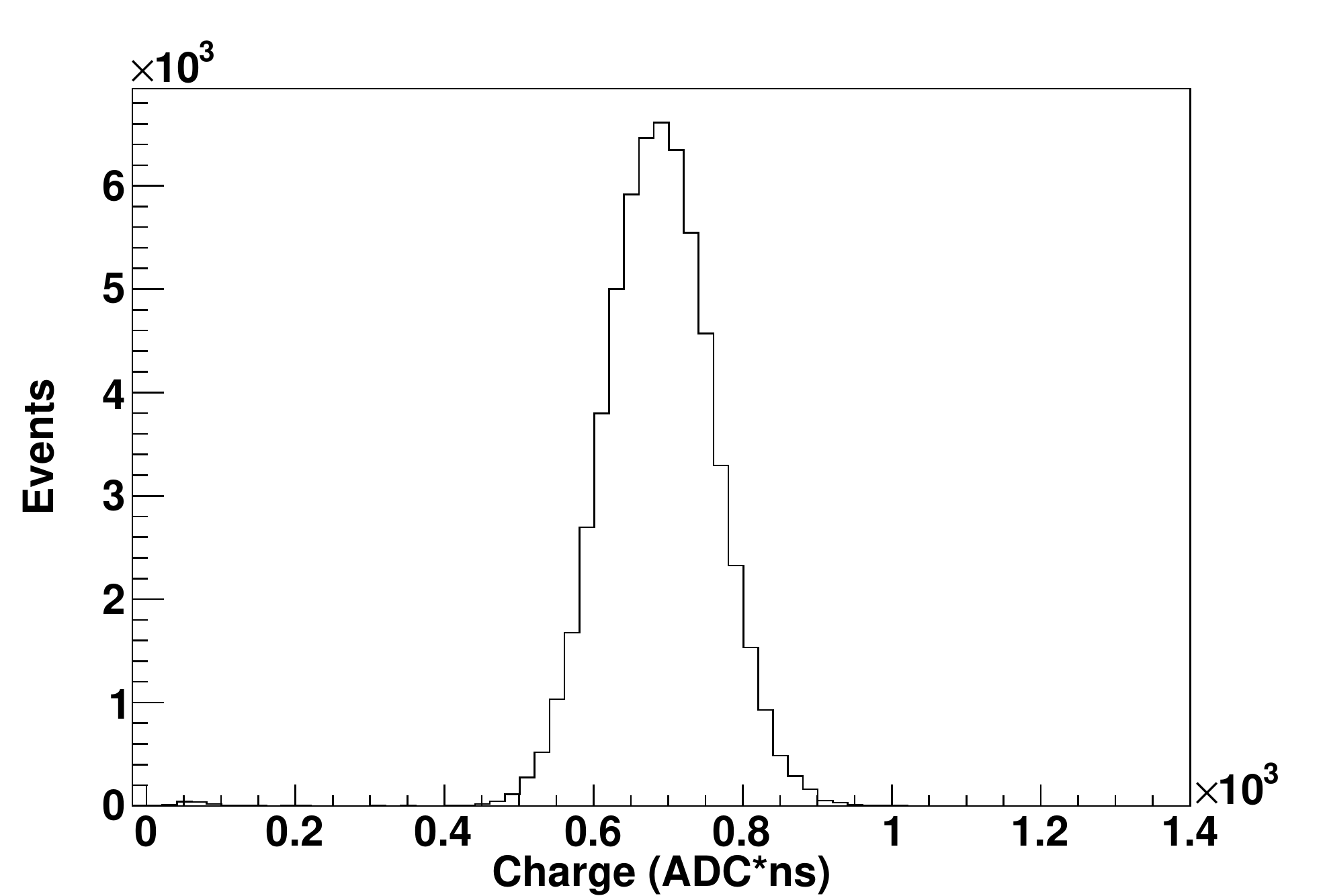} 
\caption{Distribution of the integrated charge seen by the 2'' PMT as a UV LED illuminating a spot 19.5 cm away from the end of the bar.} 
\label{fig:ex_charge_dist}
\end{figure}

\begin{figure}[t] 
\centering 
\includegraphics[width=0.49\textwidth]{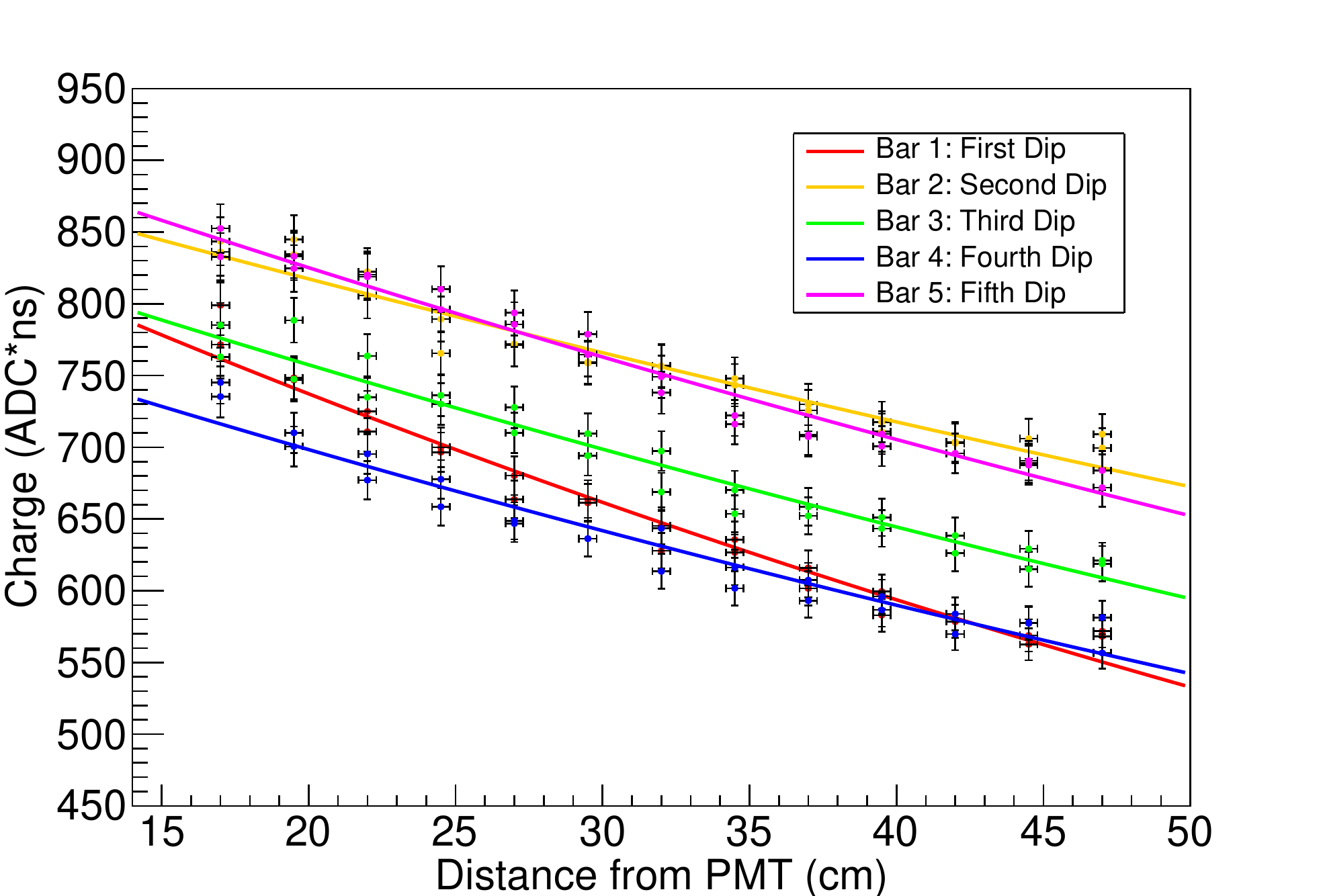} 
\includegraphics[width=0.49\textwidth]{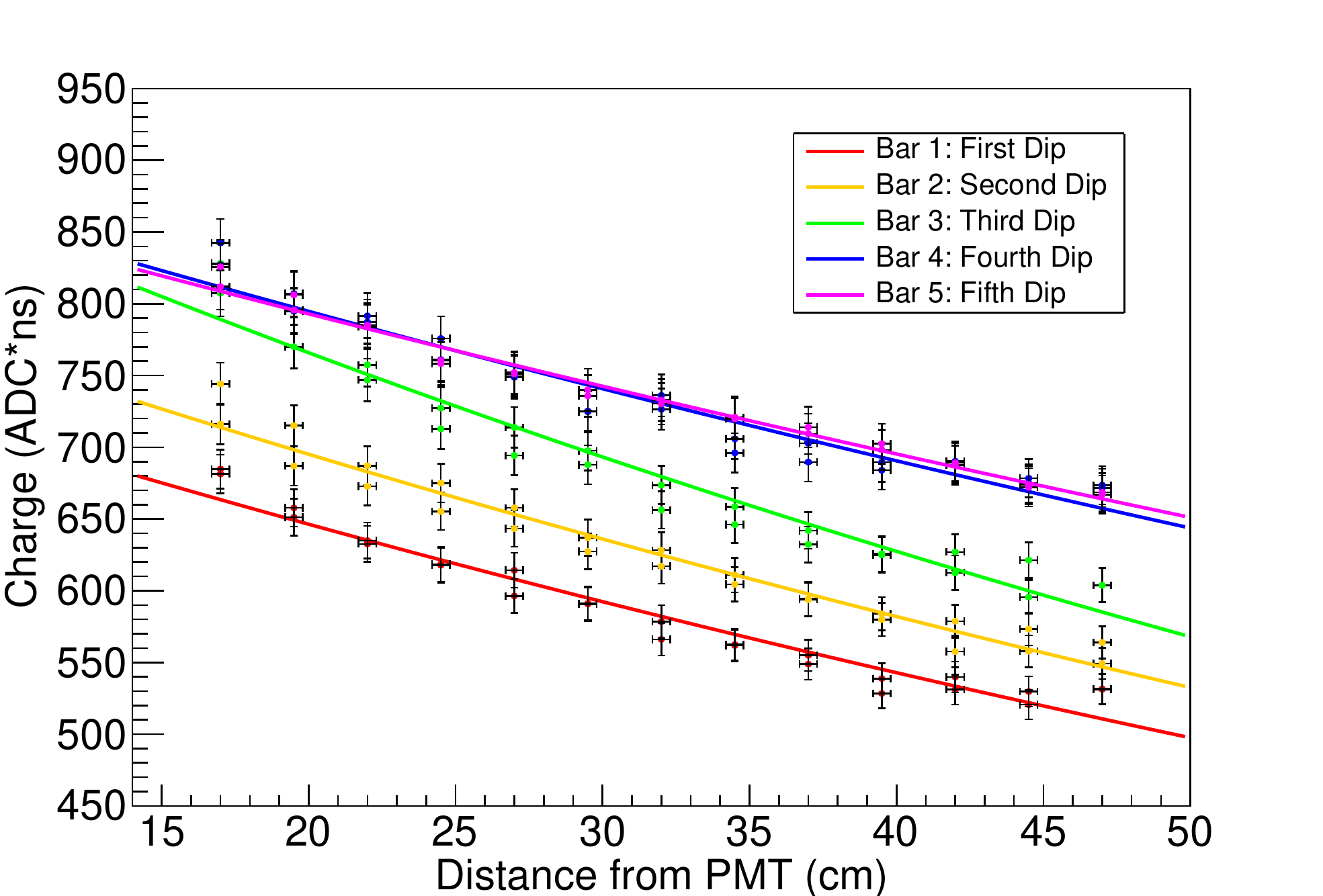} 
\caption{{\it Left:} Charge versus LED position for bars from Set 1. {\it Right:} Charge versus LED position for bars from Set 2. Note: Y axes are zero suppressed.}
\label{fig:charge_v_dist}
\end{figure}

At each measurement location along the bar, the mean integrated charge is calculated from the recorded PMT waveforms by integrating over a 60 ns window after the readout trigger.  The baseline is subtracted from this integral using an average amplitude from the pre-trigger region.  Figure~\ref{fig:ex_charge_dist} shows an example charge distribution for one of the points.  The mean charge as a function of the LED's distance from the end of the bar is used to calculate an attenuation length for each bar, found by fitting an exponential to the data.  We see no evidence for a bias when measuring the attenuation length from the first or second pass of the LED, so both sets of data are fit simultaneously.  Figure~\ref{fig:charge_v_dist} shows the set of charge measurements along the length of the bars in addition to the fitted functions for both Set 1 and 2.  

With 60,000 waveforms collected for each measurement, the total uncertainty on each measurement is expected to have a negligible contribution from the statistical uncertainty and be dominated almost entirely by the systematic uncertainty.  The repeatability of the measurement was quantified by gathering multiple data sets on a subset of four bars: bars 2 and 3 from set 1, and bars 2 and 3 from set 2. Each bar was placed in the measurement apparatus where the LED was scanned along the length of the bar, just as in the attenuation length measurements. After each scan, all power supplies, amplifiers, and signal generators were power cycled. The procedure was repeated four times per bar. 

\begin{figure}[H]
\centering 
\includegraphics[scale=0.45]{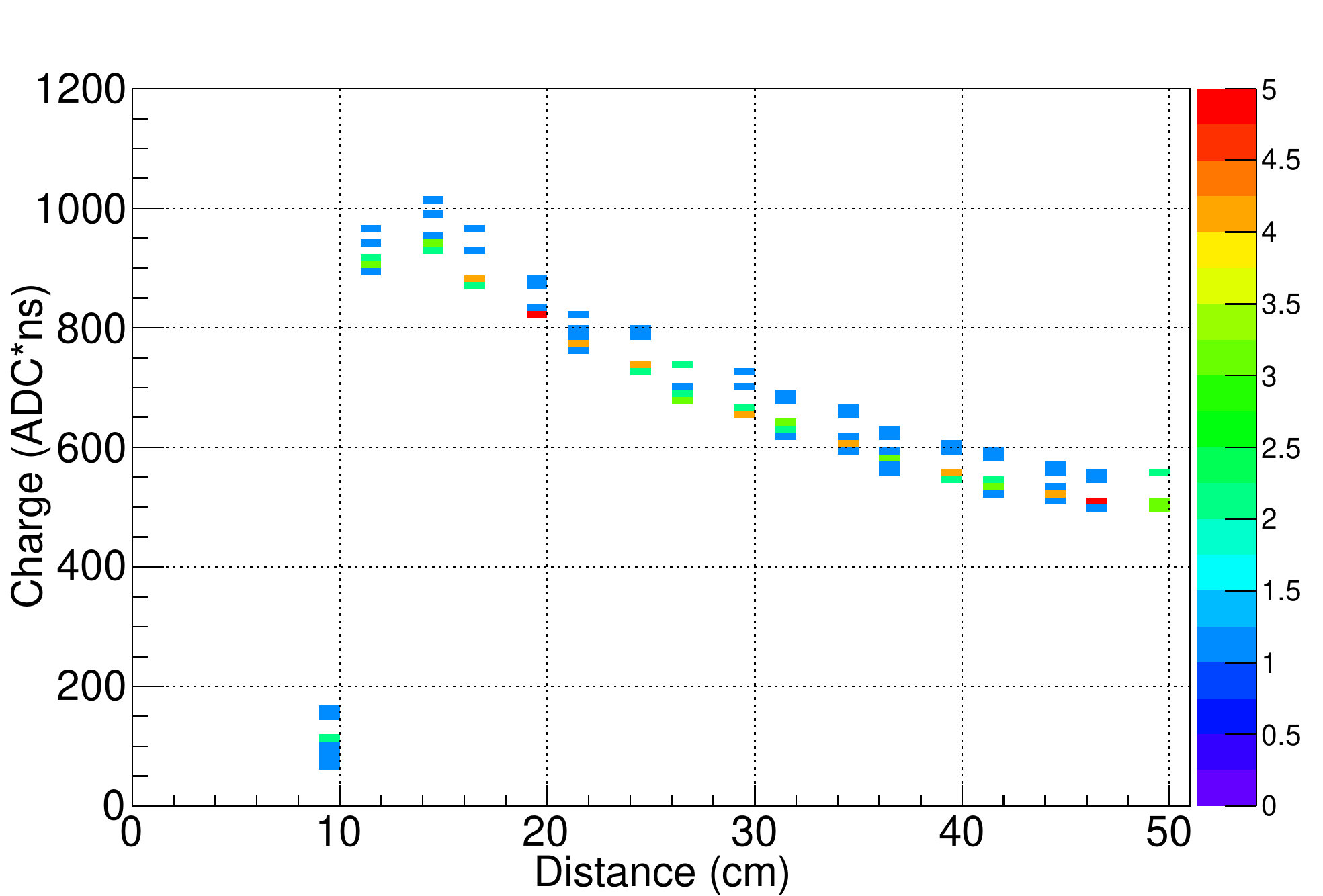} 
\caption{Charge measurements at each distance in 4 attenuation scans of bar 3, set 2.} 
\label{fig:density}
\end{figure}

An example of a repeated data set is shown in figure~\ref{fig:density}.  This bar (bar 3, set 2) exhibits charge deficits at points near the PMT caused by measurements made too close to the coating meniscus. Some of the light from the LED is projected onto a coated surface, and some onto an uncoated surface.  To mitigate this effect, all measurements at the closest distance were dropped from the analysis and cuts were placed on the event count and event charge (400 ADC$\cdot$ns).  

\begin{figure}[H] 
\centering 
\includegraphics[scale=0.45]{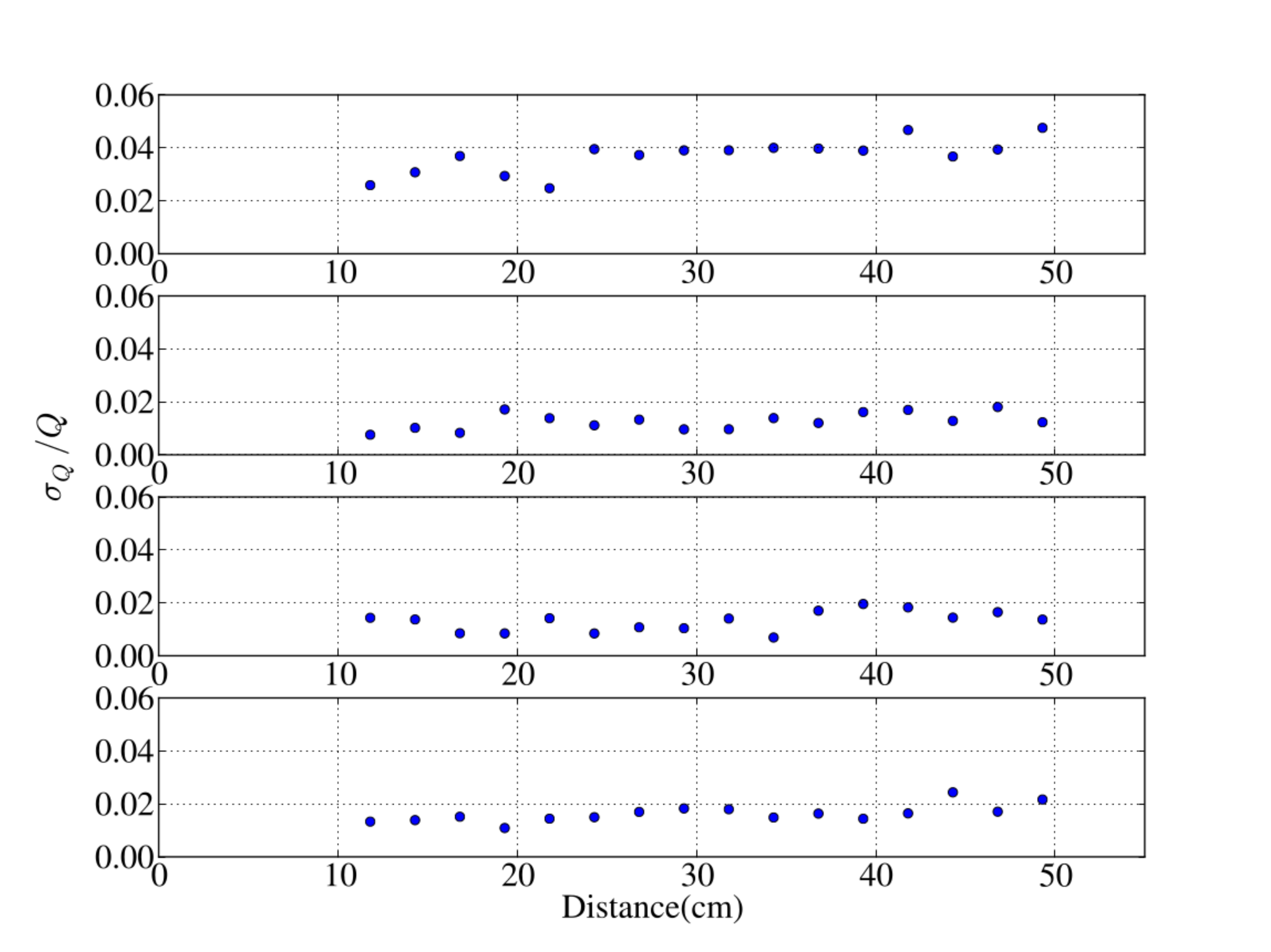} 
\caption{Fractional standard deviations of charge measurements as a function of the distance from the PMT for each bar measured. \textit{From top: set 2, bar 3; set 1, bar 2; set 2, bar 2; and set 1, bar 3}}
\label{fig:std_all}
\end{figure}

For each of the four bars, the fractional standard deviation ($\frac{\sigma_Q}{\mu_Q}$) was calculated at each position and is plotted as a function of the distance from the PMT in Figure~\ref{fig:std_all}.  The average is then taken over all distances along each bar, giving the four fractional standard deviations summarized in Table~\ref{tab:charge_stdev}.

\begin{table}
\centering
\begin{tabular}{|c|c|c|}
\hline 
Set & Bar & $\sigma_Q/\mu_Q$ \\ 
\hline 
1 & 2 & 1.28\% \\ 
\hline 
1 & 3 & 1.63\% \\ 
\hline 
2 & 2 & 1.31\% \\ 
\hline 
2 & 3 & 3.70\% \\ 
\hline 
\end{tabular}
\caption{Percent standard deviation in measured charge for four bars averaged over all distances.} 
\label{tab:charge_stdev}
\end{table}

Bar 3 from set 2 appears to have an unusually large charge variance, though it is consistent at each position. 
The average of the fractional standard deviations of these 4 bars is used as an estimate of the systematic uncertainty in each data point for our dark box measurements. The estimate for the fractional systematic uncertainty in the average charge is $1.98$\%.

\begin{figure}[t] 
\centering 
\includegraphics[scale=0.45]{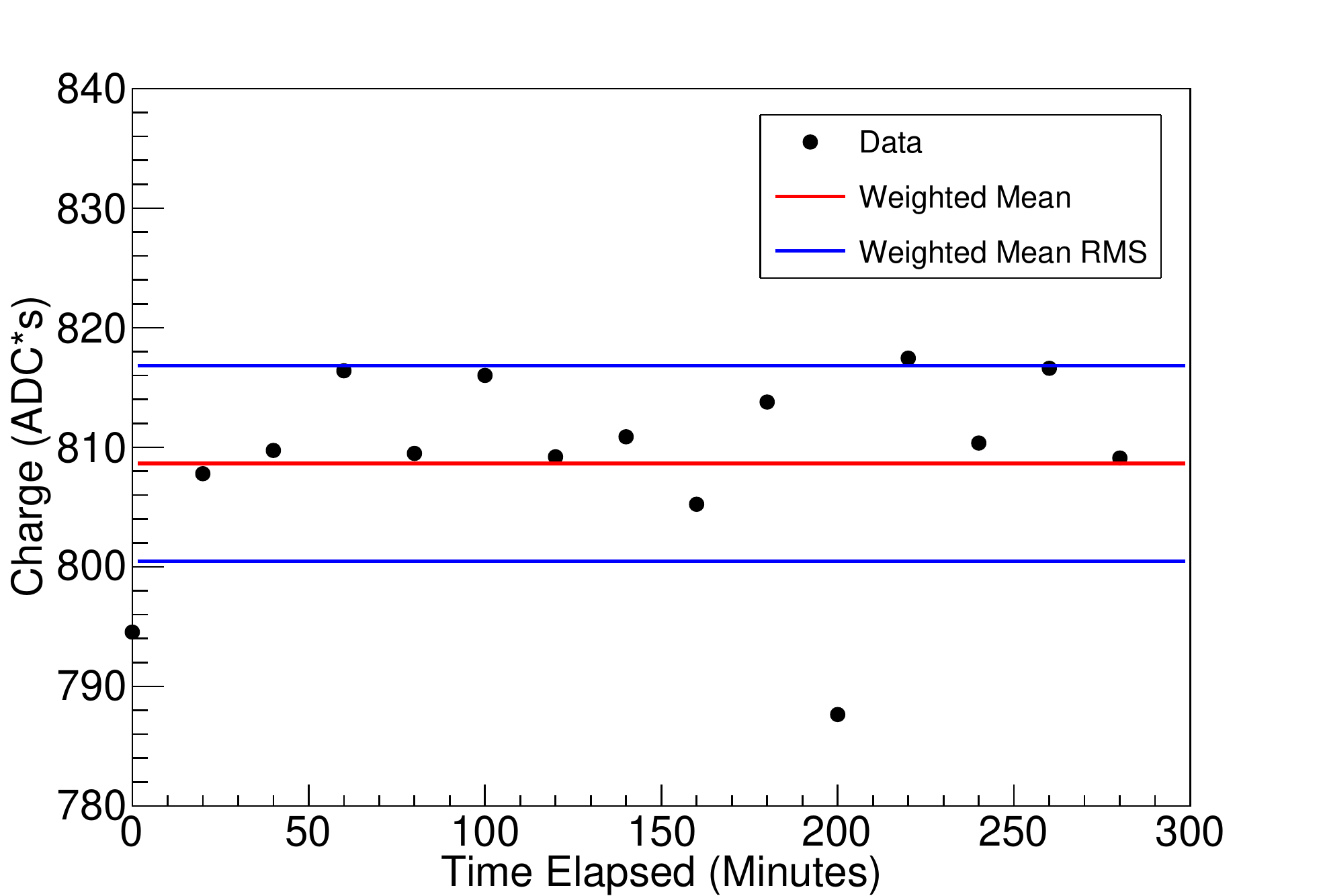} 
\caption{The mean charge response of the PMT for many repeated measurements on bar 3, set 1
  with the position of the UV LED kept constant at 21.8 cm.  Measurements were taken over 5 hours to quantify the stability of the setup. Note: Y axis is zero suppressed.}
\label{fig:LED_stability}
\end{figure}

Possible sources of the systematic uncertainty in the charge measurements
include, but are not limited to, the stability of the LED, the
precision with which the LED is positioned, the drift in the PMT gain, and the non-uniformity of
the TPB coating.  The method described above cannot differentiate between
these sources.  Therefore, a test was performed that quantifies
the uncertainty due to the stability of the LED output and PMT gain.  A coated bar was illuminated by the LED at 21.8 cm from the PMT for a
period of 4 hours and 20 minutes with the ATS9870 digitizer input voltage range set to $\pm1$V. 
Every 20 minutes $60,000$ waveforms were acquired. 
The mean of each acquisition was calculated
and is plotted in Figure~\ref{fig:LED_stability} as a function of time
elapsed.  This plot, unlike the others, displays error bars for the
statistical uncertainty only, which is given by
$\frac{\sigma_q}{\sqrt{n}}$, where $\sigma_q$ is the 
root-mean-square (RMS) of the charge distribution, and $n$ is the number of
waveforms recorded.    The spread in the measurements is much larger
than the statistical uncertainty. Therefore, the RMS about the
weighted mean can be used as the systematic error on each of the
measurements.   This comes out to a fractional systematic uncertainty
of $\sigma_{fp}=1.01\%$.  While this is a large portion of the total
systematic uncertainty, the drift of the LED and PMT cannot account for all of the
uncertainty in the charge measurements.

The best fit and the uncertainties coming from the fit for the
attenuation length and PMT charge response extrapolated to 0 cm are
given for each bar in Table~\ref{tab:air_data}.  Note that
bar 1 of set 2 was damaged at the tip facing the PMT. Although this
should not affect the attenuation length, it may affect the brightness
to an unknown extent.  The weighted mean attenuation length for sets 1
and 2 combined is $\mu=112.2\pm16.1$ cm. The weighted mean
extrapolated zero constant is $\mu=880.5\pm62.6~$ ADC$\cdot$ns.  These
values are a marked improvement over the values from previous bars
in~\cite{Baptista:2012bf}. 

\begin{table}
\centering
\begin{tabular}{| l | c | c | c |}
 \hline                       
  \multicolumn{4}{|c|}{Set 1} \\
  \hline                  
  Bar & $\lambda$ (cm) & $I_0$ (ADC$\cdot$ns) & $\frac{\chi^2}{NDF}$ \\
\hline
  1 & $92.3\pm3.6$& $915.5\pm13.0$ &$20.8/24$
  \\
  2 & $153.7\pm10.0$ & $930.1\pm13.1$ & $14.8/24$ \\
  3 & $123.7\pm6.5$ & $890.3\pm12.5$ & $16.8/24$ \\
  4 & $118.5\pm6.0$ & $826.8\pm11.7$ & $23.5/24$ \\
  5 & $127.5\pm6.8$ & $965.3\pm13.5$ & $11.8/24$ \\
  \hline  
  $\mu$ & $110.3$ & $901.1$&\\
  $\sigma$ & $17.5$ & $48.7$&\\

  \hline
\end{tabular}
\begin{tabular}{| l | c | c | c |}
  \hline                       
  \multicolumn{4}{|c|}{Set 2} \\
  \hline
  Bar & $\lambda$ (cm) & $I_0$ (ADC$\cdot$ns) & $\frac{\chi^2}{NDF}$ \\
\hline
  1 & $114.6\pm5.6$ & $769.6\pm10.9$ & $21.4/24$\\
  2 & $112.6\pm5.4$ & $830.2\pm11.7$ & $16.7/24$\\
  3 & $100.3\pm4.3$ & $934.9\pm13.4$ & $29.7/24$\\
  4 & $142.3\pm8.6$ & $914.6\pm13.0$ & $18.5/24$\\
  5 & $152.1\pm9.7$ & $904.4\pm12.6$ & $3.9/24$\\
  \hline
  $\mu$ & $114.2$ & $861.5$ & \\  
  $\sigma$ & $12.7$ & $65.5$ & \\
  \hline
\end{tabular}
\caption{Fit results of the measured PMT charge response
  as a function of UV LED position for bars produced as a part of Set 1, left, and Set 2, right.  The data is fit using an exponential function from which an attenuation length ($\lambda$) and the extrapolated
  PMT charge response at 0 cm from the end of the bar ($I_0$) are estimated.
  } 
\label{tab:air_data}
\end{table}

\subsection{Orientation of the Bars}
\label{subsec:orientation_air}
At the end of the dip coating process, the light guides are hung to dry after being withdrawn from the coating solution. Although the evaporation occurs quite quickly thanks to the volatility of toluene, it is possible that the cohesion of the liquid on the surface of the bar, in combination with gravity, drags the solution down the bar; the effect of which would be a coating thickness gradient along the length of the bar.  Depending on the thickness scales involved and on the penetration depth of the UV light encountering the bar, such a gradient may cause a variation in light output as a function of excitation position that is separate from the effect of optical attenuation.  In addition, surface losses occurring as light is guided are also expected to depend on the coating thickness, although this dependence is neglected in the following calculations.
 
The bars are measured with their uncoated ends facing the PMT, so the ``bottoms'' of these bars (bottom in the hang-drying orientation) are facing away from the PMT.  This would imply that any thickness gradient present would increase in the direction of increasing distance. If the penetration depth of the incident UV light is greater than the minimum thickness, then this could cause higher light output at the far end of the bar, and improve the measured optical attenuation length. If the bar were reversed end to end and measured again, the effect should be reversed, causing a higher light output at the near end of the bar, and a concomitant reduction of the measured optical attenuation length.  

The transmission of 286 nm UV light through a 0.100'' thick sheet is quoted by the manufacturer to be around 60.2\%\cite{UTRANspec}.  Therefore, we expect that the penetration of 286 nm light into our coating, $\mathcal{O}$(0.1''), is deep enough to produce a coating gradient effect.  (We also note because the UV transmission is quoted to be near zero for 260 nm and less, which implies that for liquid argon 128 nm scintillation light, we do not expect the effect to be seen for bars in liquid argon as will be discussed in Section~\ref{sec:model}.)  To study the gradient effect in air, we compare measurements of the attenuation length of a bar oriented both ``forwards'' and ``backwards'' in our setup.   Here, we assume that fractional loss due to any scattering at the end of the bar facing the PMT when oriented backwards is the same for each LED position.

\begin{figure}[t] 
\centering 
\includegraphics[scale=0.50]{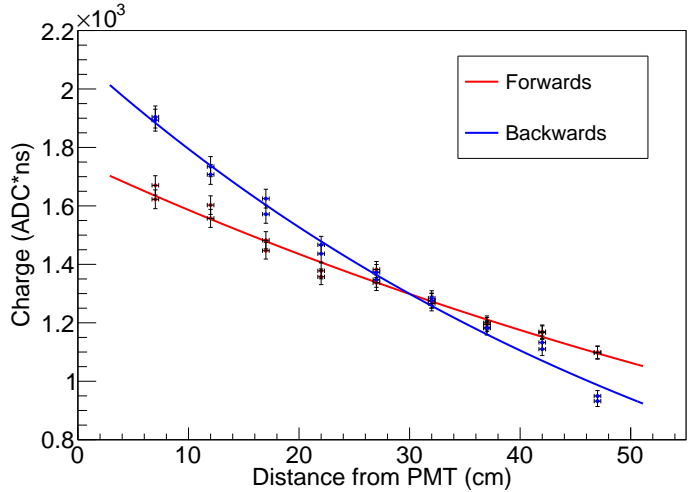} 
\caption{its to the PMT charge response as a function of UV LED 
position for a bar oriented "forwards" and "backwards" in the measurement setup. The error bars are hidden within the data points. The change in measured attenuation length from forward to backward orientation was $\Delta \lambda = -38.1 \pm 2.7$\% cm. Similarly, the change in EZB was $\Delta I_0 = 20.4 \pm 1.9$\% ADC$\cdot$ns. Note, Y axis is zero suppressed.}
\label{fig:301-A-1_direction}
\end{figure}

The results are shown in Figure~\ref{fig:301-A-1_direction}. The difference in the measured attenuation length and PMT charge response extrapolated to 0 cm between the bar oriented forwards and backwards in our setup are given as $\Delta \lambda$ and $\Delta I_{0}$, respectively.   We measure the ``backwards'' attenuation length to be $61.9\pm1.4$ cm, which is a decrease in the attenuation length of 38.1 $\pm$ 2.7\% for a bar oriented backwards with respect to a bar oriented forwards ($100.0\pm3.6$ cm).  This is consistent with a gradient in our coating thickness, which increases in the direction of gravity as the bars are hung to dry.  

\subsection{Effects of Repeated Dip Coating}
\label{sec:air:repeat_dips}

If the production of these bars were to be scaled up, the dipping procedure would need to be optimized for more efficient use of the coating solution. To this end, dipping multiple batches of bars in the same solution would drastically reduce fluid consumption, while speeding the production process.   
As can be seen in Table~\ref{tab:air_data}, 9 out of 10 of the bars in sets 1 and 2 were measured to have an attenuation length that surpassed a benchmark value of >1 m. However, we do see a substantial variation in the measured attenuation lengths.  

We tested whether this variation could be attributed  to a contamination or degradation of our coating solution when reused for multiple bars.  The responses of the two sets of 5 bars discussed above are plotted with ordering information in Figure 9.  We find no evidence of monotonic changes of the PMT charge response extrapolated to 0 cm (the "extrapolated zero brightness" or EZB) or attenuation length up to 5 consecutive dips.

\begin{figure}[t] 
\centering 
\includegraphics[scale=0.80]{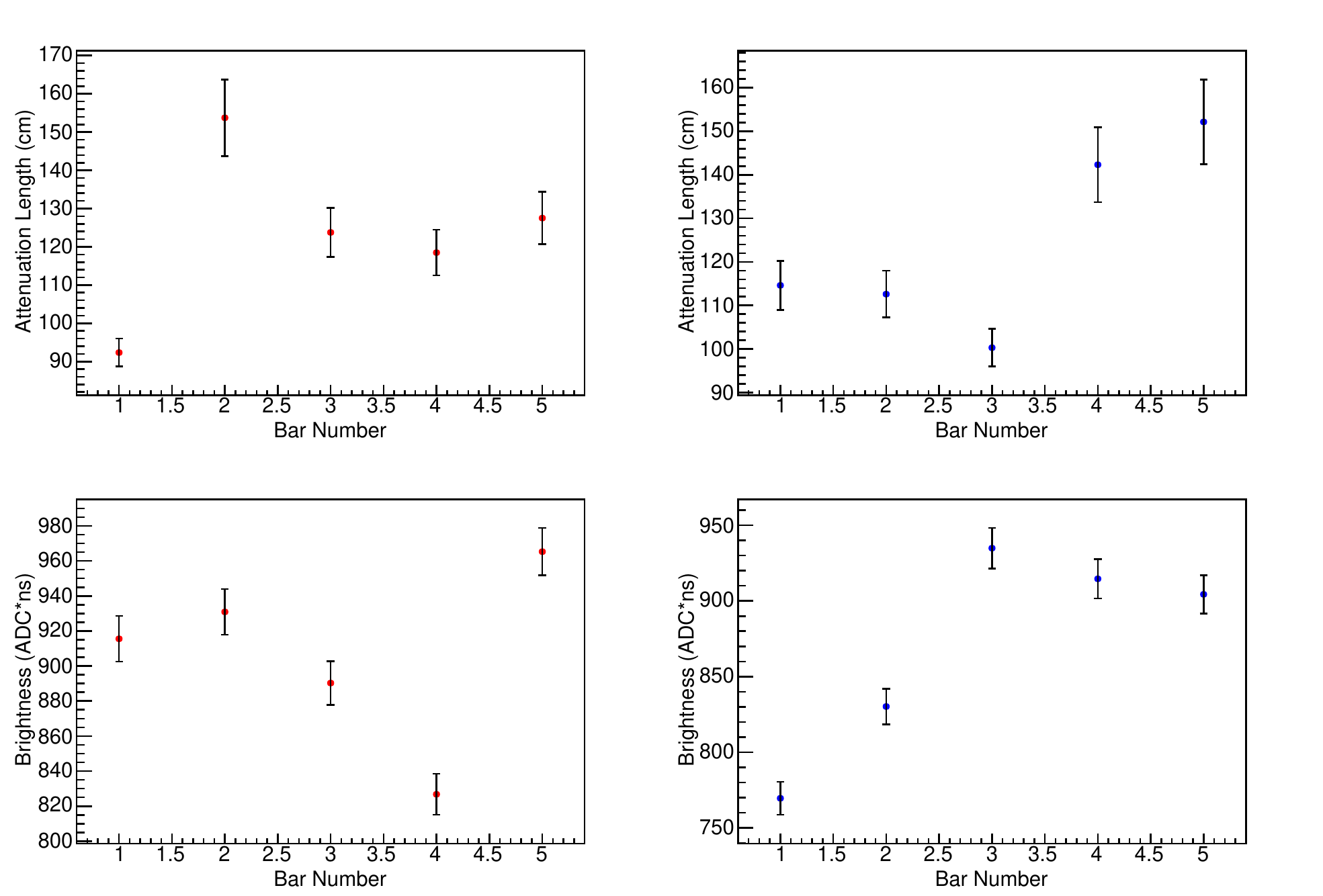} 
\caption{Attenuation lengths and extrapolated zero-length brightness as a function of dip order for bar sets 1 and 2. \textit{Upper left}: Set 1, attenuation length. \textit{Upper right}: Set 2, attenuation length. \textit{Lower left}: Set 1, extrapolated zero brightness. \textit{Lower right}: Set 2, extrapolated zero brightness. Note: Y axes are zero suppressed.}
\label{fig:dip_order_plot}
\end{figure}

\input{TallBoAttenuationMeasurements2.tex}

\section{Conclusion}
\label{sec:conclusions}
Improvements to the production of TPB-coated light guides suitable for LAr scintillation light detection in neutrino LArTPCs were described, in which a novel TPB solution is applied to a cast UVT acrylic bar using a hand dip-coating method. The attenuation lengths of TPB-coated light guides were studied in detail with a 286 nm LED in the lab.  The measured attenuation lengths extend above 100 cm, which is a notable improvement over previous work and achieves an important benchmark in the development of meter-scale light guides for future LArTPC light collection systems currently being designed.  The attenuation behavior of these light guides were then connected to measurements made in liquid argon using a simple model.  This illustrates how measurements in air can be used to quantitatively predict the attenuation behavior of light guides in liquid argon for future large scale LArTPCs.

\section*{Acknowledgments}

This material is based upon work supported by the National Science Foundation under Grant No. NSF-PHY-1205175.  We thank Stephen Pordes for allowing us access to the Bo cryostat and its associated cryogenic system.  We are very grateful to Bill Miner, Ron Davis, Terry Tope and the other technicians who have assisted us at the Proton Assembly Building, Fermilab.
TW also gratefully acknowledges the support provided by the Pappalardo Fellowship Program at MIT and the Intensity Frontier Fellowship at Fermi National Laboratory.

\newpage

\bibliographystyle{JHEP}

\bibliography{references}

\end{document}

%% file: TallBoAttenuationMeasurements2.tex
\section{Attenuation Length Measurements in Liquid Argon}
\label{sec:attlar}


While the measurements in air serve as a useful and convenient benchmark, the performance of the light guide bars in high purity liquid argon is what really determines their suitability as a light collection system for future LArTPCs.  Therefore, the attenuation behavior of two of the light guide bars were also measured in a high purity liquid argon test stand called ``TallBo''.

TallBo is a cylindrical, vacuum-insulated cryostat 22'' in diameter and 70'' tall located at Fermilab's Proton Assembly Building.  Its cryogenics system includes a series of regenerable filters shared with the Materials Test Stand described in~\cite{Andrews:2009zza}.  Prior to each fill, the cryostat is evacuated for several days with a turbo pump to remove contaminants.  TallBo is then filled with high-purity liquid argon with residual impurities of H$_2$O, O$_2$, and N$_2$ at the sub-ppm level.   

The pressure inside TallBo is kept at $\sim10$ psig by a liquid-nitrogen-cooled condenser tower, which re-condenses argon vapor into the liquid. When run with the condenser, the cryostat operates as a closed system, as has been demonstrated in Refs~\cite{Jones:2013bca, Jones:2013mfa}.  The liquid level is measured using a capacitive level monitor and for these studies was maintained a few inches below the lid of the cryostat.

\subsection{Experiment Setup}

The TPB-coated acrylic light guide bar that is being measured is oriented vertically and read out on one end by a 2'' cryogenic PMT (Hamamatsu R7725MOD).  It is held in place at each end by a pair of set screws and is initially installed a fraction of an inch above the PMT.  When the setup is submerged in LAr, the PMT's buoyancy brings it in contact with the end of the bar.  To prevent stray light from reaching the PMT, it is enclosed in an aluminum housing with a small window on the top through which the light guide bar protrudes and small gaps on the bottom to allow argon to flow and for cabling.

The light guide is then mounted in front of five $^{210}$Po disk sources, which are distributed at equal 4'' intervals along the length of the bar.  Each disk source is enclosed in an identical aluminum holder, which has a 0.2'' diameter recess that is 0.3'' deep.  The distance of each source to the light guide bar is 0.9'', so that the direct light emitted from the center of the source illuminates a 0.6'' diameter spot on the light guide bar.  

Five SiPMs (SensL Micro-FB-60035) are installed on the opposite side of the light guide directly across from each disk source and are used for triggering.  The 5.3 MeV alphas from the $^{210}$Po disk sources deposit their energy in the LAr and produce uniform, point-like sources of 128 nm scintillation light that illuminate five different positions on the bar.  After the VUV light is wavelength-shifted to visible light in the coating on the surface of the bar, a portion of it will be guided toward the end of the bar and collected by the 2'' PMT while another portion of it will be re-emitted in the forward direction, where it can be collected by the corresponding SiPM.  

\begin{figure}[h] 
\centering 
\includegraphics[width=0.95\textwidth]{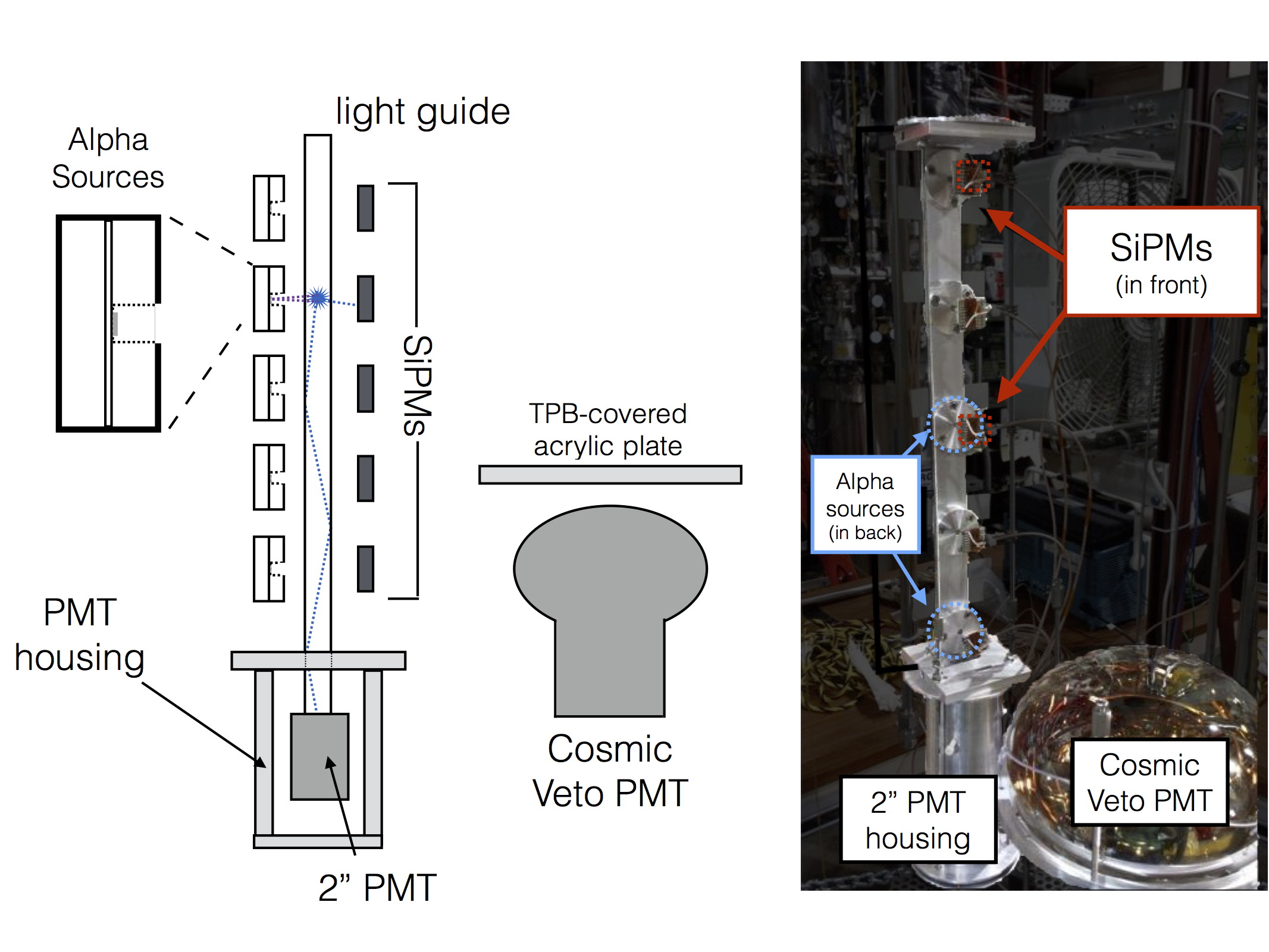} 
\caption{A diagram and photograph of the TallBo attenuation length measurement setup.  Not shown in the photograph are the TPB-coated acrylic light guide and the TPB-covered acrylic plate.  Also not shown in the diagram or the photograph is the optical fiber used to calibrate the 2'' PMT.}
\label{fig:tallbo_diagram}
\end{figure}

To reject cosmic ray backgrounds, a PMT 
and a TPB-covered acrylic plate are installed to the side of the light guide bar. 
For calibration, a quartz optical fiber directs blue light to the 2'' PMT from an external LED driven by a board with a design similar to Ref.~\cite{Conrad:2015xta}.  A schematic of the setup is given in Figure~\ref{fig:tallbo_diagram}.

\subsection{Calibrations}

The setup involves several components that require calibration.  This includes the 2'' PMT collecting photons at the end of the bars, the SiPMs monitoring the sources, and the sources themselves. The calibration of the 2'' PMT and SiPMs was performed \textit{in situ}, in conjunction with the measurements in the Tall Bo cryostat.  The calibration of the alpha sources was performed in a separate apparatus.

\subsubsection{2'' PMT Calibration}
\label{subsec:tallbo_pmt_calibration}

The gain of the 2'' PMT is measured \textit{in situ} using the LED, driver board, and optical fiber setup described above.  The driver board is used to pulse the LED and trigger the readout of the 2'' PMT.  The LED output is tuned so that on average the  2'' PMT photocathode emits <1 photoelectron for each LED pulse.  The 2'' PMT single photoelectron (SPE) distribution is then obtained by integrating the baseline-subtracted waveforms in a fixed window around the trigger.

The 2'' PMT SPE charge distribution is modeled by Eq.~\ref{spe_equation}, which is a sum of 3 terms.  The first two terms are weighted by the Poisson probability of detecting zero photoelectrons $e^{-\lambda}$ and describe a Gaussian pedestal with mean $\mu_{ped}$ and variance $\sigma^2_{ped}$ and exponential noise with decay constant $\tau_{noise}$.  The relative weighting between these two terms is given by $(1-\beta)$ and $\beta$, respectively.  The third term describes charge due to the emission of one or more photoelectrons from the photocathode, which is modeled as a series of Gaussians with means and variances given in terms of the single photoelectron mean $\mu_{SPE}$ and variance $\sigma^2_{SPE}$ and with areas determined by Poisson statistics.

\begin{figure}[h]
\centering
\includegraphics[width=0.9\textwidth]{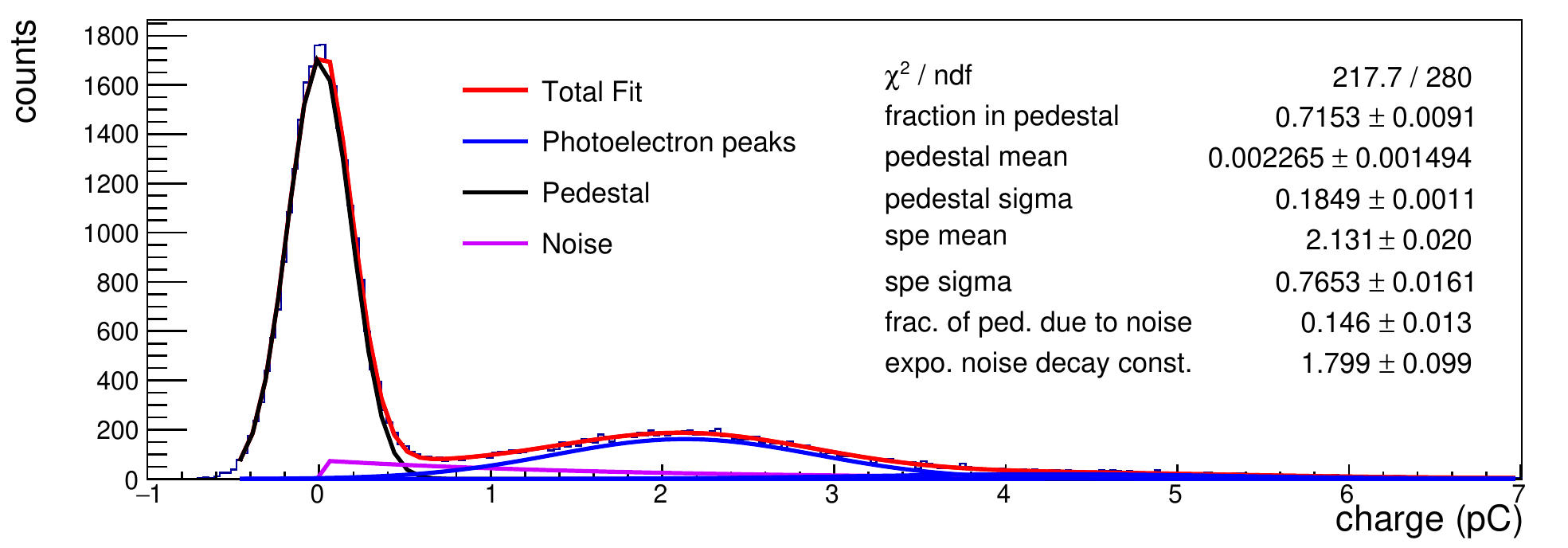} 
\caption{An example fit to the 2'' PMT single photoelectron (SPE) charge distribution.}
\label{fig:ex_pmt_spe}
\end{figure}

The parameters $\beta$, $\lambda$, $\sigma_{ped}^2$, $\mu_{ped}$, $\tau_{noise}$,  $\sigma_{spe}^2$, and $\mu_{spe}$ are extracted from a fit to the SPE charge distribution $N(x)$ with integral $N_{tot}$ and bin width $w$ given by

\begin{align}
N(x) & = N_{ped}(x) + N_{noise}(x) + N_{pe}(x) \nonumber \\
& = N_{tot}e^{-\lambda}(1-\beta)\frac{w}{\sigma_{ped}}\sqrt{\frac{1}{2\pi}}e^{-\frac{(x-\mu_{ped})^2}{2\sigma_{ped}^2}}\nonumber\\
& + N_{tot}e^{-\lambda}\beta\theta(x-\mu_{ped}) \frac{w}{\tau_{noise}} e^{-\frac{x-\mu_{ped}}{\tau_{noise}}}\nonumber \\
& + N_{tot}\sum_{n=1}^{N}\frac{w}{\sqrt{n}\sigma_{spe}}\frac{\lambda^n e^{-\lambda}}{n!}\sqrt{\frac{1}{2\pi}}e^{-\frac{(x-n\mu_{spe})^2}{2n\sigma_{spe}^2}},
\label{spe_equation}
\end{align}

\noindent where $\theta(x)$ is the Heaviside step function.  An example SPE charge distribution fit is given in Figure~\ref{fig:ex_pmt_spe}.

\begin{figure}[h]
\centering
\includegraphics[width=0.9\textwidth]{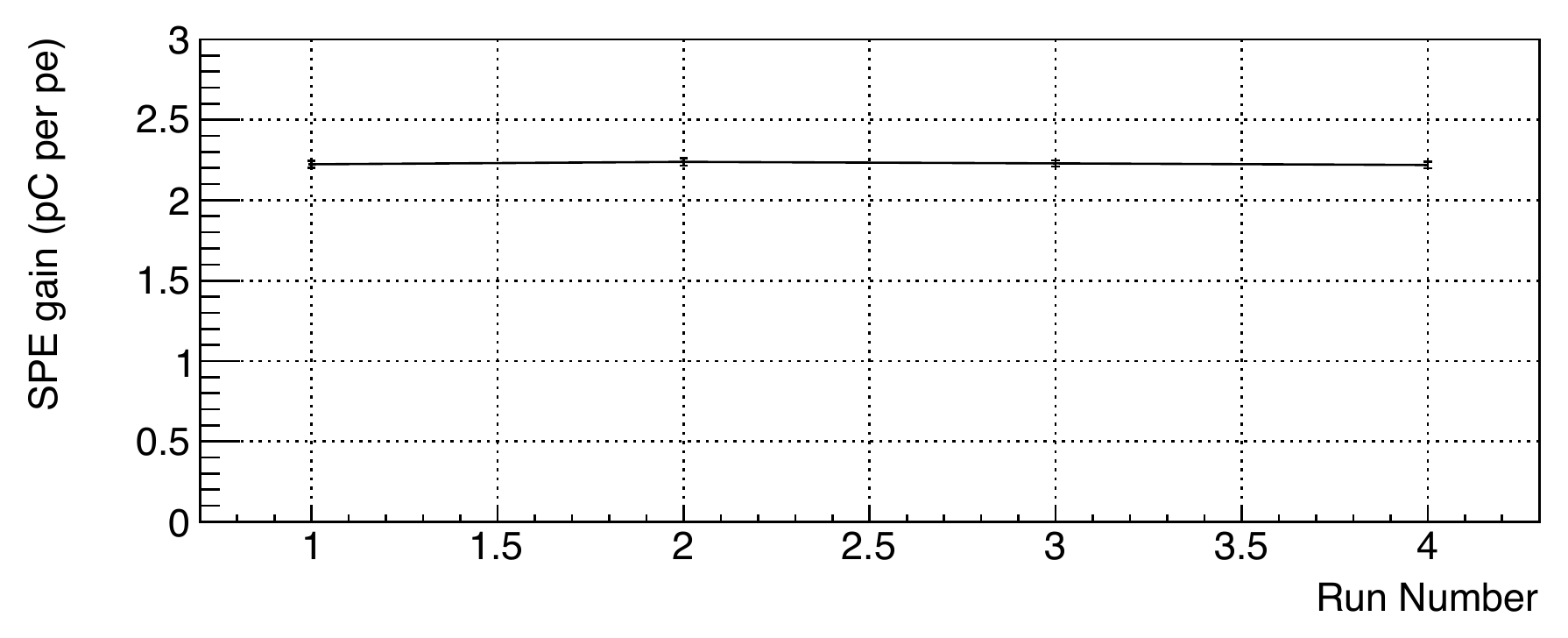}
\caption{Fitted values of the single photoelectron (SPE) mean taken over the course of one attenuation length measurement runs.}
\label{fig:ex_spe_stability}
\end{figure}

The SPE mean $\mu_{spe}$ and variance $\sigma_{spe}^2$ are measured periodically throughout data taking in order to monitor drift in the PMT gain. Figure~\ref{fig:ex_spe_stability} shows the SPE mean extracted from calibration data taken over one attenuation length run in between the five SiPM-triggered data sets.  The 2'' PMT is observed to be stable for not just the run shown in the figure, but for all runs.  The standard deviation in the measured SPE mean is used as a systematic uncertainty in the measurement and contributes about one percent.

\subsubsection{SiPM Calibration}
\label{subsec:sipm_calibration}

\begin{figure}[h]
\centering
\includegraphics[width=0.75\textwidth]{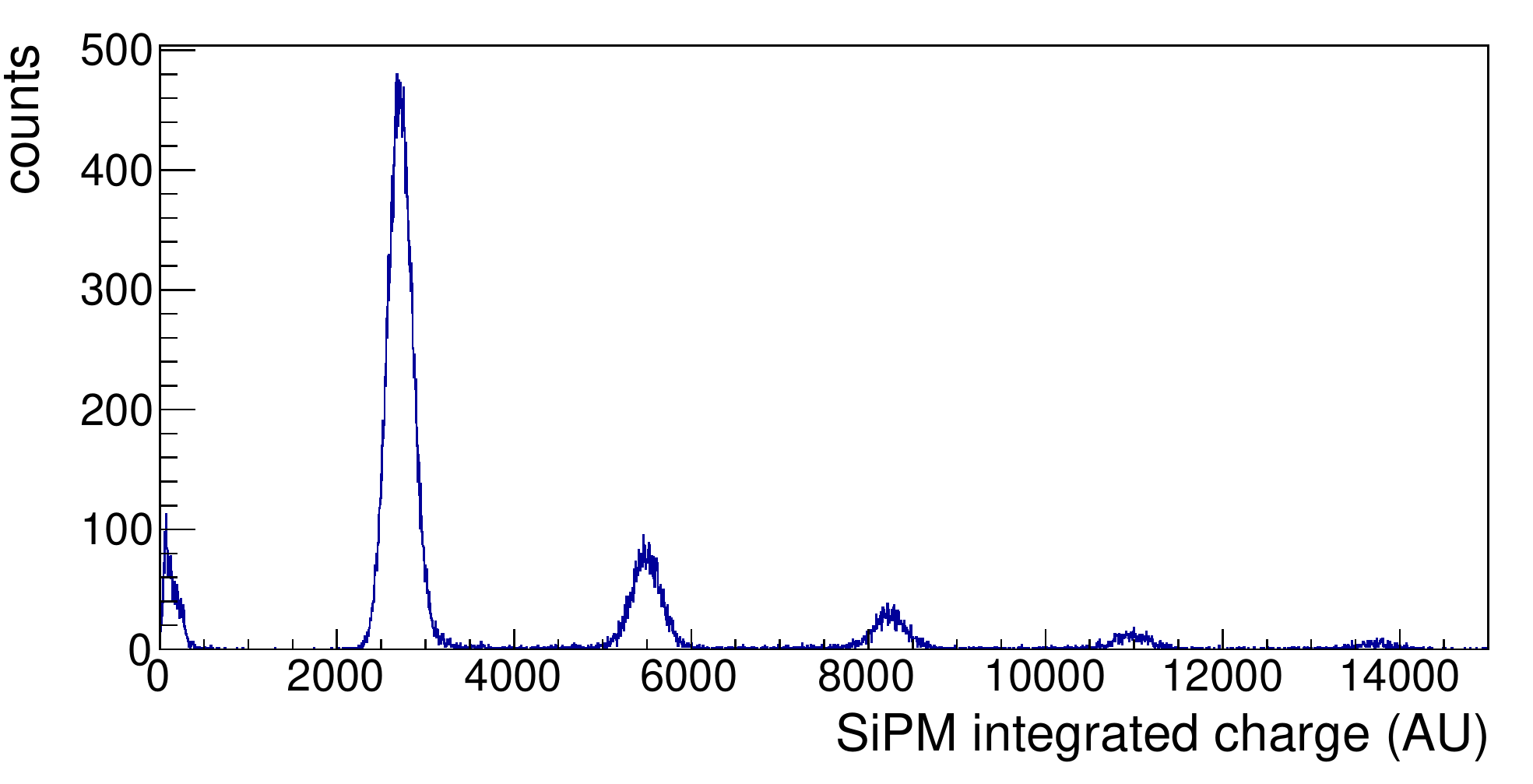}
\caption{Example SiPM integrated charge distribution. The mean of the first peak (at 2735) is used to determine the single photoelectron response of the SiPM.}
\label{fig:ex_sipm}
\end{figure}

The gain of the SiPMs can also be determined for each data set by extracting the SPE mean $\mu_{SPE}$.  SiPMs are composed of large arrays of Geiger-mode avalanche photodiodes.  
At low light levels the individual PE peaks in the integrated charge spectra are very well-defined, because they correspond to an integer number of micro-cells firing.
A simple pulse-finding algorithm is used to integrate the SiPM waveforms (see Section~\ref{sec:tallbo_analysis}).  
Figure~\ref{fig:ex_sipm} shows an example histogram of the charge distribution seen by a SiPM.  
In the histogram, the clear separation between photoelectron peaks can be seen.  
The single photoelectron response is determined by taking the mean of the first peak beyond the low-charge events coming from noise.

\subsubsection{Alpha Source Calibration}
\label{subsec:alpha_calibration}

\begin{figure}[h]  
\centering  
\includegraphics[width=0.95\textwidth]{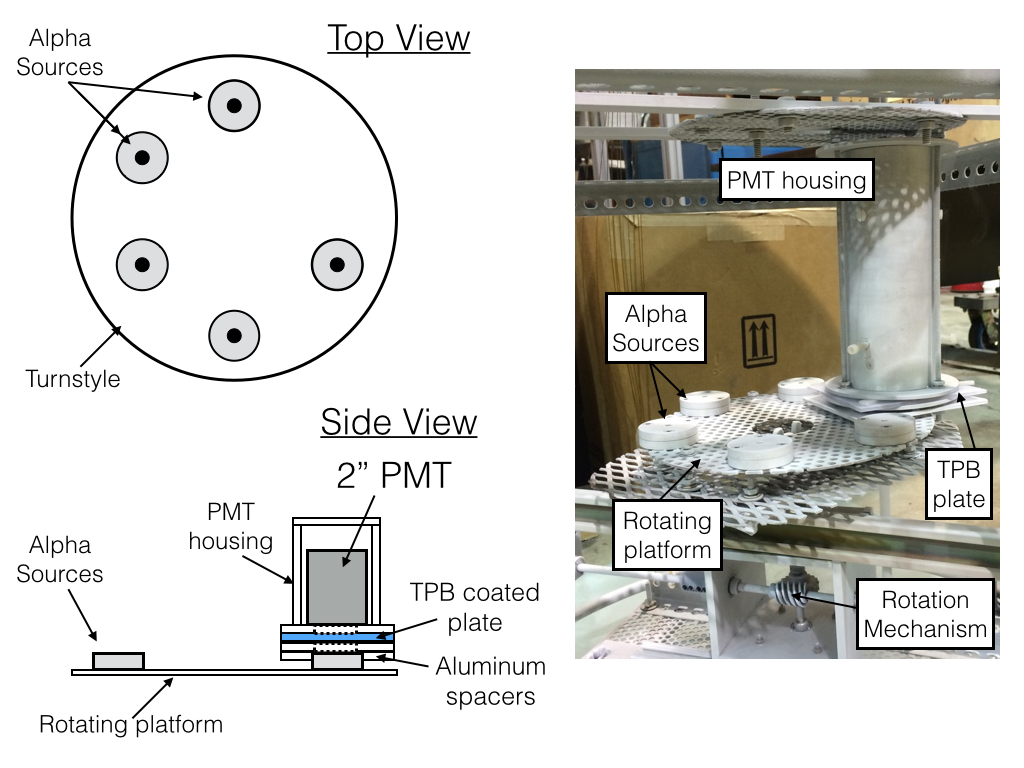} 
\caption{The alpha source calibration setup. \textit{Left}: A top and side view diagram of the setup.  \textit{Right}: A photo of the setup (covered in frost after just having been removed from the liquid argon).}
\label{fig:stella_diagram}
\end{figure}

To measure the relative amount of LAr scintillation light produced by each of the $^{210}$Po sources, the disk sources and their aluminum holders were installed at 60$^{\circ}$ intervals about the axis of a rotating platform in a wide-mouth dewar filled with research-grade liquid argon.  A series of worm wheels and gear boxes allowed the five sources and an unoccupied position on the platform to be rotated beneath a fixed PMT assembly \textit{in situ}.  The PMT assembly consisted of a 2'' cryogenic PMT (Hamamatsu R7725mod) in an aluminum housing and a 3'' $\times$ 3'' TPB-covered acrylic plate sandwiched between two 0.125'' thick aluminum spacer rings, each with a 1.5'' inner diameter and a 3.5'' outer diameter.  There were small gaps at the top and bottom of the housing for cabling and to allow argon to flow.  To align the fixed PMT assembly with the rotating sources, an additional aluminum spacer was mounted on the bottom of the assembly with a 2.125'' guide hole matching the diameter of the source holders.  The entire PMT assembly was then lowered directly onto each source holder (or the unoccupied position) using a long threaded rod which penetrated the lid of the dewar.  A schematic of the setup is given in Figure~\ref{fig:stella_diagram}.

Since this dewar is open to atmosphere, care must be taken to mitigate the introduction of impurities such as O$_2$ and N$_2$ during the fill, which can affect the amount of scintillation light collected.  This is accomplished by first purging the vessel with argon gas to push out the initial volume of air in the dewar and then filling the vessel with liquid argon through a pipe extending to the bottom of the dewar.  The vessel is allowed to fill quickly until the PMT assembly is fully submerged, at which point the fill rate is slowed but not stopped during data taking in order to maintain a positive overpressure in the dewar.

\begin{figure}[h] 
\centering
\includegraphics[width=0.95\textwidth]{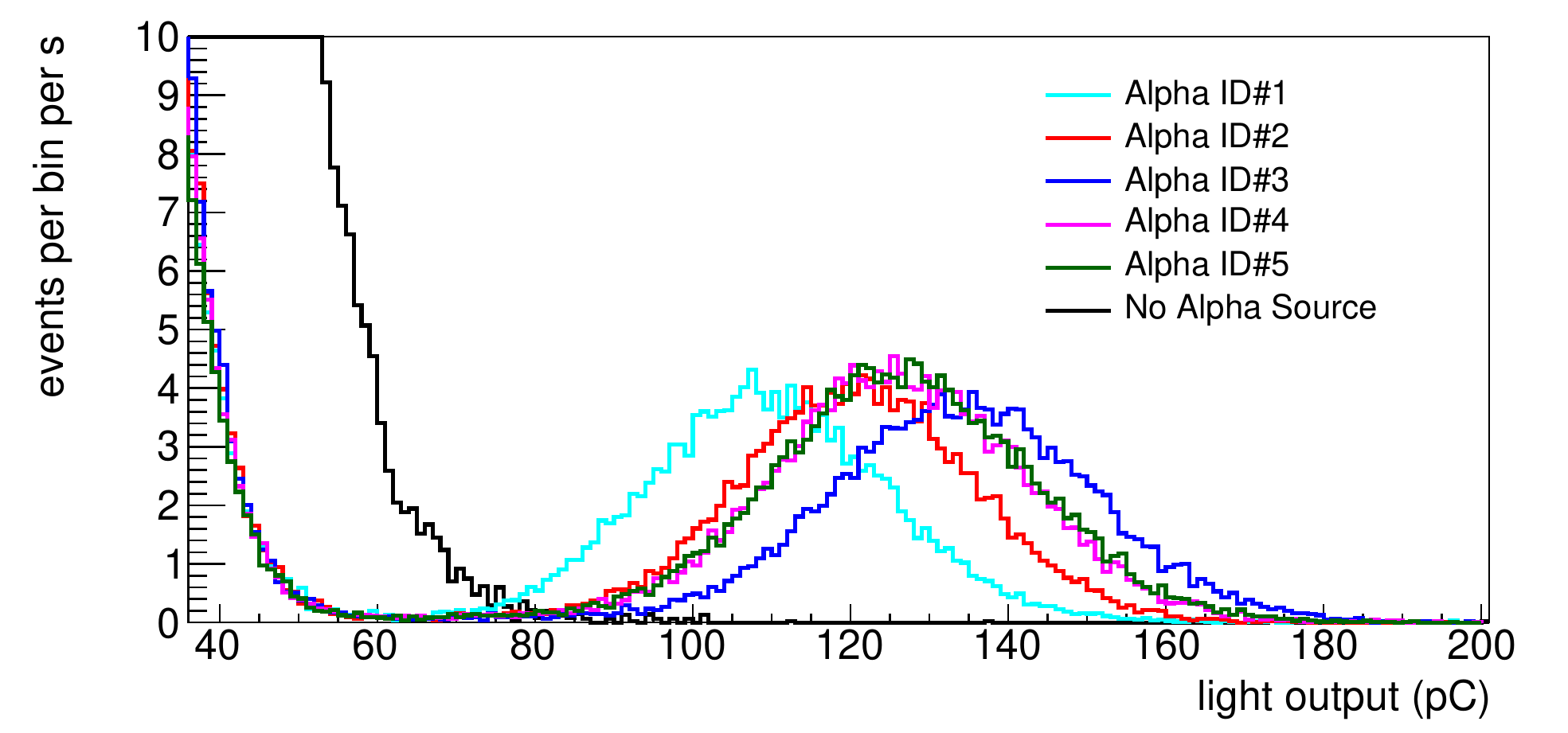} 
\caption{PMT charge distributions corresponding to each of the five $^{210}$Po alpha sources as well as a control data sample with no source.  Further details are given in the text.}
\label{fig:stella_peaks}
\end{figure}

Digitized waveforms were collected using an oscilloscope triggering on the 2'' PMT for the five $^{210}$Po alpha sources and the unoccupied position on the platform four separate times over a period of $\sim5$ hours. Figure~\ref{fig:stella_peaks} shows the integrated charge distributions normalized to the total run time corresponding to each alpha source and to the position on the platform with no alpha source.  The alpha peaks are well separated from the background and exhibit a clear variation over the five sources.  As expected, no peak is present in the data acquired at the unoccupied position though in this case the background is larger because the TPB plate is not shadowed by a source holder.

\begin{figure}[h] 
\centering
\includegraphics[width=0.95\textwidth]{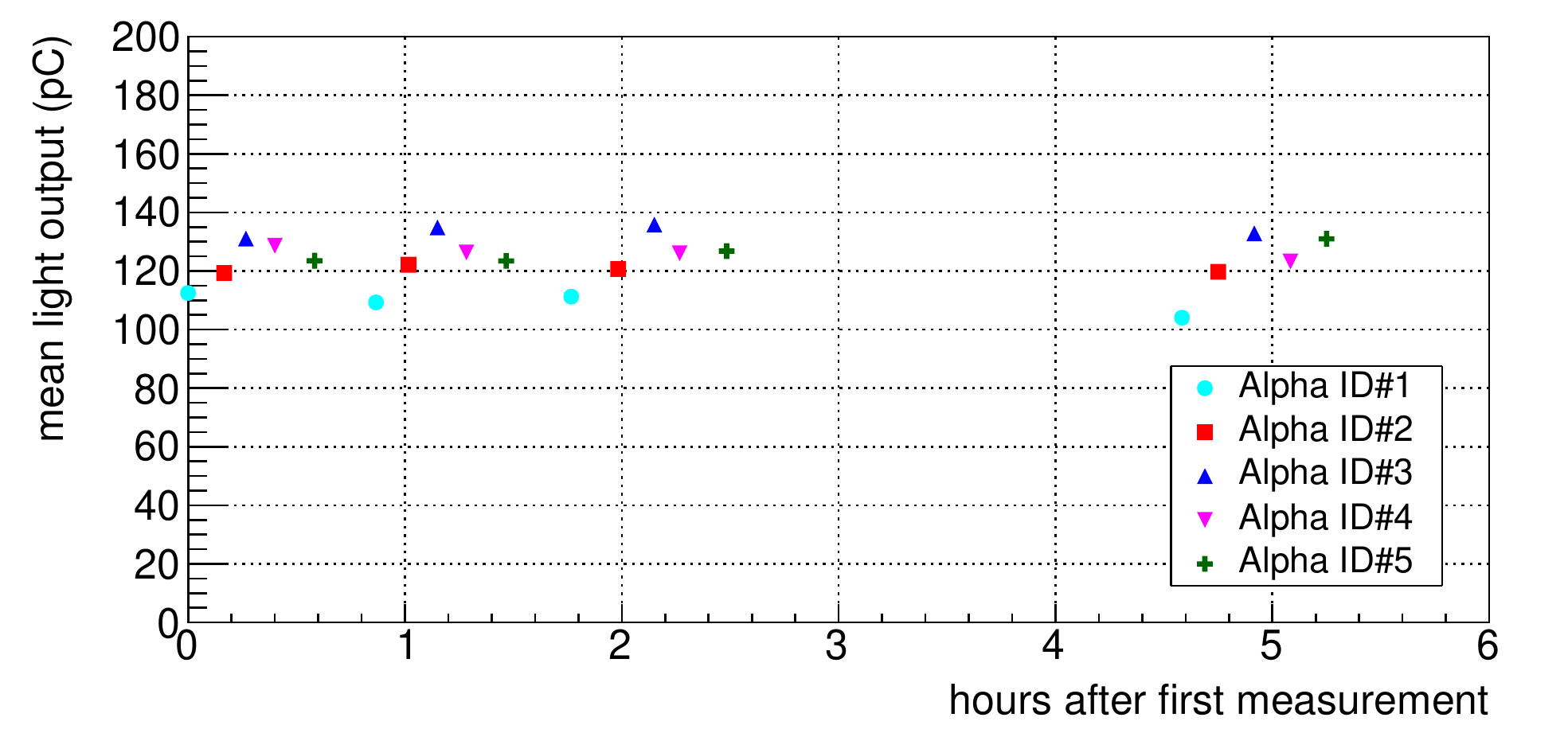}
\caption{Mean value of a Gaussian fit to the PMT charge distributions corresponding to each alpha source as a function of time.}
\label{fig:stella_mean_vs_time}
\end{figure}

We determined the systematic error in the relative mean charge of the five alpha peaks due to any time variations by fitting the four measurements of each $^{210}$Po source individually.  The mean of the Gaussian fit to the alpha peaks of each source as a function of the elapsed time since the beginning of the data taking is given in Figure~\ref{fig:stella_mean_vs_time}.  

The weighted mean of the four measurements for each alpha source is given in Table~\ref{tab:alpha_correction}.  A correction factor is computed for each source relative to the measured charge in the alpha peak averaged over the five sources.  
These correction factors and their associated uncertainties are then propagated to the analysis of the light guide bar attenuation length data.

\begin{table}[h]
\centering
\begin{tabular}{ l  c c }
  \multirow{2}{*}{Alpha ID\#} & \multirow{2}{*}{error-weighted mean} &
  \multirow{2}{*}{ratio to combined mean}\\
\\
\hline

  1 & $109.2\pm3.7$ & $0.89\pm0.03$\\ 
  2 & $120.7\pm1.2$ & $0.98\pm0.01$\\ 
  3 & $133.8\pm2.0$ & $1.09\pm0.02$\\ 
  4 & $125.9\pm2.3$ & $1.02\pm0.02$\\ 
  5 & $126.3\pm3.6$ & $1.03\pm0.03$\\ 
  \hline
combined mean & $123.2$ & 1.0 \\
\end{tabular}
\caption{Error-weighted mean charge of the four measurements of each alpha source.  A correction factor for each source is also given relative to the measured alpha peak charge averaged over the five sources.} 
\label{tab:alpha_correction}
\end{table}

\subsection{Electronics Readout}

The PMTs and SiPMs drive several meter long 50 $\Omega$ RG-316 cables, which penetrate the cryostat through a potted epoxy feedthrough.  The signals are AC coupled, and in the case of the SiPMs also amplified, before being digitized and read out by a pair of oscilloscopes.  A copy of the cosmic veto PMT and SiPM signals are also sent to NIM logic, which is used to build a trigger.  The oscilloscopes are simultaneously triggered by a discriminated SiPM signal, which is put in anti-coincidence with a $6.4~\mu$s gate generated by the discriminated cosmic veto PMT signal.  An Arduino micro-controller is used to enable the anti-coincidence logic and re-synchronize the data acquisition between the two oscilloscopes every 23 triggers.  In addition, a copy of the triggered SiPM signal is sent to both oscilloscopes, so that offline analysis can determine if the oscilloscopes are reading out the same event.  


\subsection{Analysis}
\label{sec:tallbo_analysis}

First, a straightforward pulse-finding algorithm is used to build the SiPM and 2'' PMT charge spectra from their digitized waveforms.  The algorithm computes the mean and standard deviation of the baseline in a fixed pre-trigger region and then looks for a pulse which crosses a threshold of 4 standard deviations from the baseline mean.  The baseline-subtracted pulse is then integrated starting 200 ns before this threshold crossing ($t_{start}$) and ending 300 ns after the pulse falls back below threshold.  The SiPM and 2'' PMT pulse charge spectra are then converted into photoelectrons using the calibrations described in Sections~\ref{subsec:tallbo_pmt_calibration} and~\ref{subsec:sipm_calibration}.

Next, two data quality cuts were used to isolate a set of good events for analysis.  One cut removed blocks of 23 events if an oscilloscope synchronization problem was found for any those events.  This was determined by applying a $\chi^2$ comparison test to the waveforms of the triggering SiPM that were copied to both oscilloscopes.  The other cut removed events for which an anomalously high baseline RMS was computed, indicating the presence of pulses in the pre-trigger region that could bias the baseline estimation.   The synchronization cut removed about 3.5\% of the events from the data set. The synchronization and baseline cuts together remove about 6\% of the events.

A series of three cuts were then used to select a clean set of 2'' PMT pulses detected in coincidence with a SiPM signal corresponding to an alpha decay from one of the five $^{210}$Po sources.  The first of these cuts required that the difference between the $t_{start}$ of the 2'' PMT and the triggering SiPM be between 0 -- 800 ns to reduce random coincidences.  The second cut required that the the number of PEs detected by the triggering SiPM be greater than any other SiPM.  This reduced the crosstalk between source/SiPM pairs, when a SiPM triggers on the light produced by an alpha from a neighboring $^{210}$Po source.  The last cut required the triggering SiPM to detect >1 PEs to reject events caused by dark pulses in the SiPM.  Depending on which SiPM/source is used to trigger, the selection efficiency of the cuts ranges between 16\% to 32\%.

Finally, the calibrated 2'' PMT spectrum for the selected events is fit to a multi-photoelectron distribution $MPE(x)$ with a floating normalization, whose form is given by the third term in Equation~\ref{spe_equation} but with $\mu_{SPE}\equiv1$ and $N=40$:
\begin{equation}
MPE(x) = C\sum_{n=1}^{N=40}
\frac{e^{-\lambda}\lambda^{n}}{n!}e^{-\frac{(x-n)^2}{2n\sigma_{spe}^2}}.
\label{eq:mpe_func}
\end{equation}

\begin{figure}[h]
\centering
\includegraphics[width=0.95\textwidth]{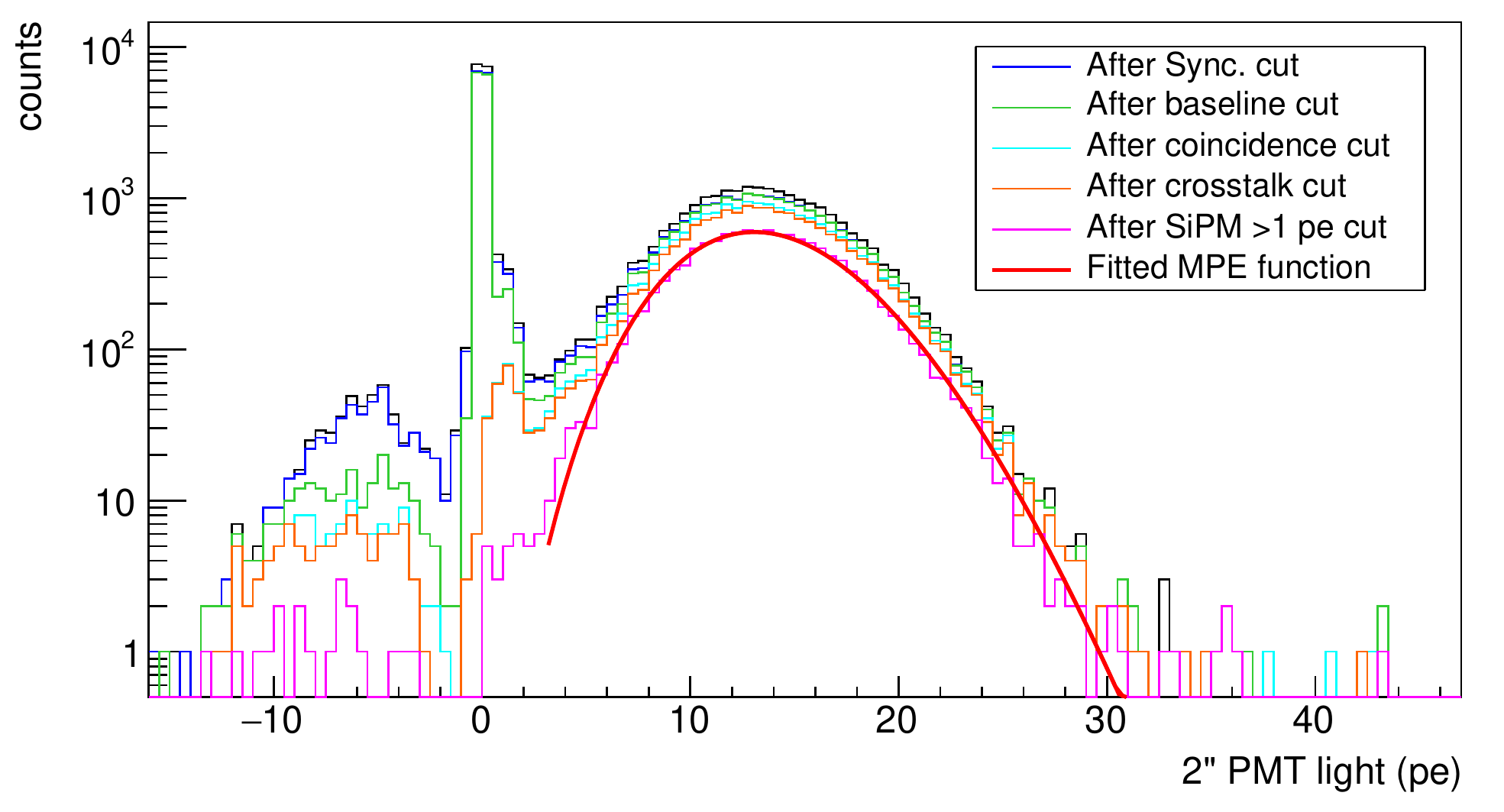} 
\caption{2'' PMT charge spectrum calibrated in terms of photoelectrons with various cuts applied as well as an example multi-photoelectron fit to the data.}
\label{fig:ex_pedist_wcuts}
\end{figure}

\noindent Figure~\ref{fig:ex_pedist_wcuts} shows a calibrated 2'' PMT spectrum with all of
the different cuts applied as well as an example $MPE$ fit to the data.

The average number of photoelectrons $\lambda$ extracted from the $MPE$ fit to the data was then corrected according to Table~\ref{tab:alpha_correction} to account for variations in the $^{210}$Po alpha sources and plotted as a function of distance from the 2'' PMT along the bar.  Data were taken for two different light guide bars, referred to as ``bar A'' and ``bar B'' below.  An attenuation length measurement consisted of 5 sets of data, corresponding to $\sim40,000$ triggers from each of the 5 SiPMs before cuts.  In order to study the systematic effect of the ordering of the alpha sources along the bar, bar A was measured twice with two different alpha source orderings.  The source ordering for bar B, run 1 was the same as for bar A, run 1, whereas in bar A, run 2 each source was shifted one position further from the PMT except for the last source which was moved to the position closest to the PMT. 

\begin{figure}[h]
\centering
\includegraphics[width=0.95\textwidth]{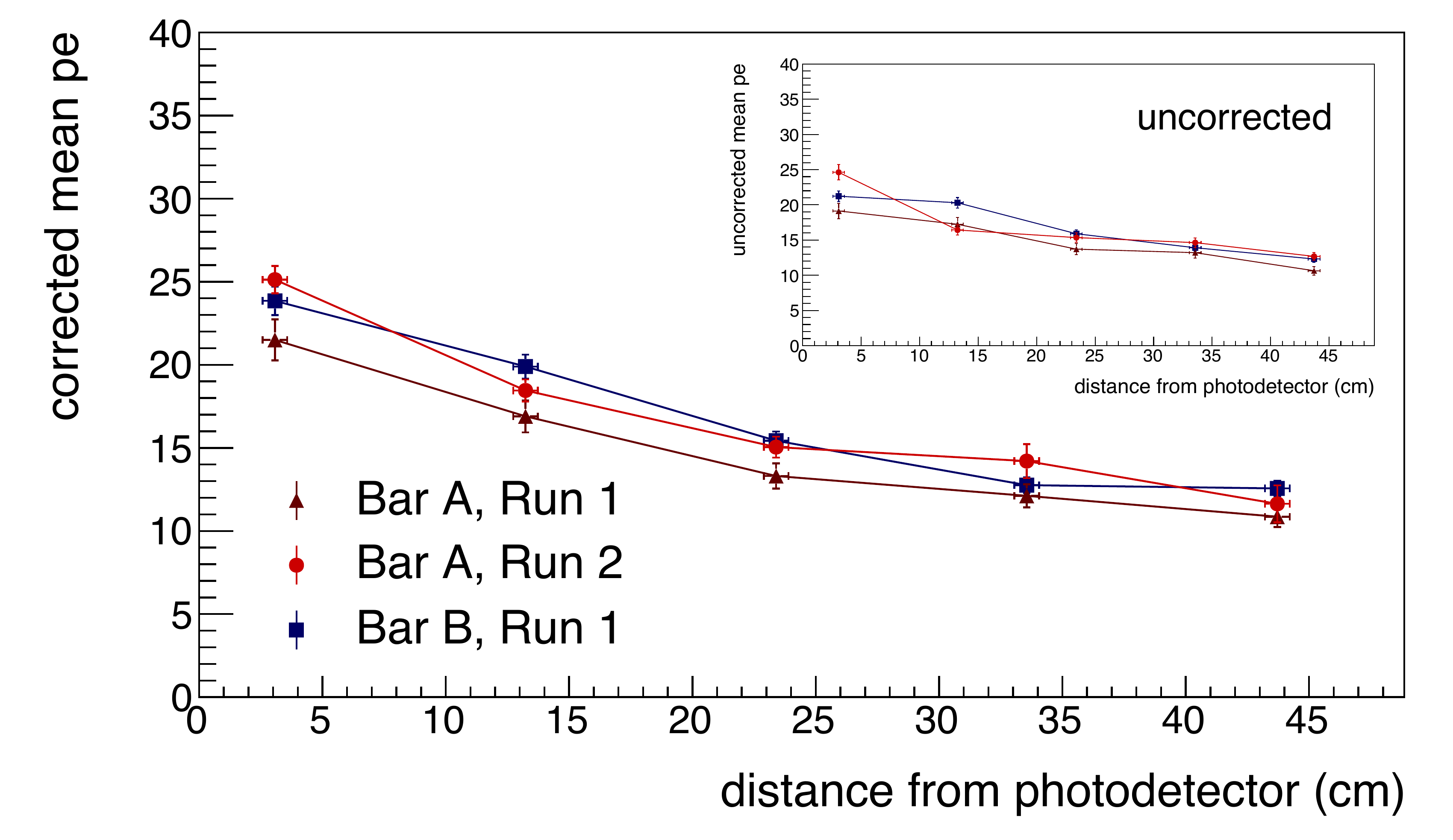} 
\caption{Corrected mean number of photoelectrons collected by the 2'' PMT as a function of the distance of the source from the PMT along the bar after correcting for the relative light output of the alpha sources.  The inset in the upper right shows the data without the alpha source correction applied. }
\label{fig:lar_bar_response}
\end{figure}

As can be seen in Figure~\ref{fig:lar_bar_response}, the variation in the light output of the sources influences the shape of the measured curves.   Before correcting for the relative light output of the sources, the shape of the curve for bar A is different for runs 1 and 2.  Furthermore, the shape of the  curves for bar A, run 1 and bar B, which were collected with the same ordering of alpha sources, are similar.  However, after the alpha source correction factors are applied, the curves for bar A, run 1 and bar A, run 2 are similar. This suggests that we have calibrated out the relative light output of the sources using the external measurement described in Section~\ref{subsec:alpha_calibration}.

\subsection{Attenuation Length in Liquid Argon}

\begin{figure}[h]
\centering
\includegraphics[width=0.95\textwidth]{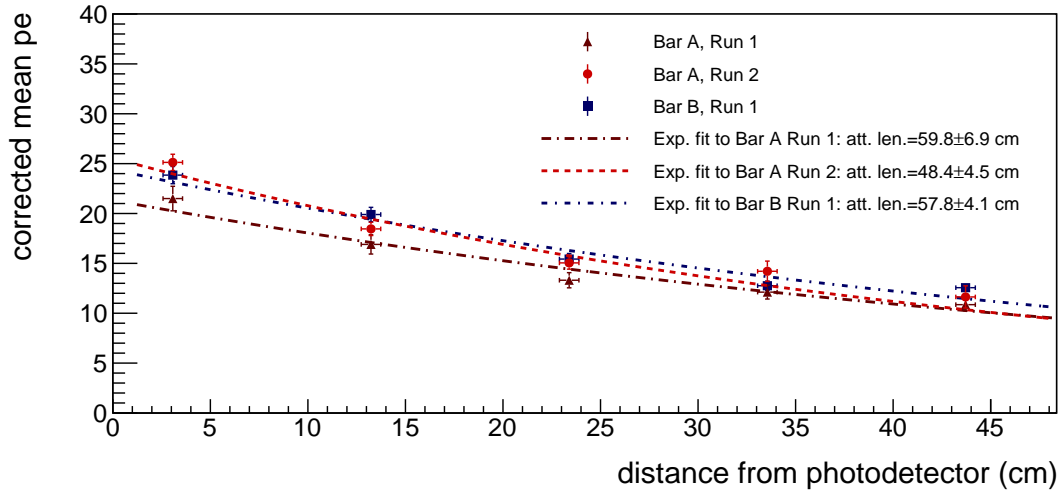}
\caption{Exponential fits to the corrected mean number of photoelectrons collected by the 2'' PMT as a function of the distance of the source from the PMT along the bar.}
\label{fig:lar_bar_expofit}
\end{figure}

Following previous conventions, an exponential function is fit to the light output data in order to extract an attenuation length for the tested bars.  Figure~\ref{fig:lar_bar_expofit} shows the data and the best fit exponential functions.  Table~\ref{tab:expo_fits} gives the resulting attenuation lengths from the fits along with the $\chi^2/\textrm{d.o.f}$ for each fit.  For comparison, the table also lists the attenuation lengths measured in air for both the forward and backward orientations.  Compared to earlier versions of the bars, which saw attenuation lengths around 20 cm,  these bars are a significant improvement.  However, it is clear from the figure and the $\chi^2$ values that an exponential function does not always fit the data very well.  Furthermore, it is difficult to see how to connect the attenuation lengths found in air with those found in liquid argon.  To address this issue, we describe a model that can be used to characterize the output of the bars in both air and liquid argon in Section~\ref{sec:model}.

\begin{table}[h]
\centering
\begin{tabular}{ l c c c c c }
\multirow{2}{*}{Bar} & \multirow{2}{*}{Run} &  LAr fit & LAr fit & Air  fit, forward & Air fit, backward\\
& & att. length (cm) & $\chi^2$/d.o.f. & att. length (cm) & att. length (cm) \\
\hline
A & 1 & $59.8\pm6.9$ & 4.0/3 & \multirow{2}{*}{100.0} & \multirow{2}{*}{61.9}\\ 
A & 2 & $48.4\pm4.5$ & 8.2/3  & & \\ 
B & 1 & $57.8\pm4.1$ & 12.5/3 & 70.4 & 67.0 \\  
\end{tabular}
\caption{Attenuation lengths calculated from exponential fits to the corrected mean number of photoelectrons collected by the 2'' PMT as a function of the distance of the source from the PMT along the bar measured in liquid argon.} 
\label{tab:expo_fits}
\end{table}

\subsection{Liquid Argon Purity Checks}
\label{sec:lar_purity}

Table~\ref{tab:impurity_levels} lists measurements of the N$_2$, O$_2$, and H$_2$O concentrations made either before or after each run.  In cases where the reading on the monitor was oscillating, the highest recorded value is given.  The upper limits on the measured impurity concentrations can be used to predict the possible effects of light quenching over the course of a run.  In our data we expect any such effects to be dominated by residual O$_2$ contamination since the rate constant of light quenching for O$_2$ is a factor of 5 larger than for N$_2$~\cite{WArP_nitrogen, WArP_oxygen}.

\begin{table}[h]
\centering
\begin{tabular}{ l c c c }
Run & O$_2$ (ppb) & H$_2$O (ppb) & N$_2$ (ppb) \\
\hline
Bar A, Run 1 & <400 & <25 &    <87 \\   
Bar A, Run 2 &   <130 &  <34 & <400 \\ 
Bar B, Run 1 &    <270 & <15 &  <220 \\ 
\end{tabular}
\caption{Measured concentration of impurities in the liquid argon for each data run.}
\label{tab:impurity_levels}
\end{table}

We constrain any time variation in the scintillation light yield for the runs with the highest upper limit on the O$_2$ impurity concentration by making repeated measurements.  For bar B, run 1, the entire set of attenuation measurements was repeated after $\sim$1 week.  The mean number of photoelectrons measured by the 2'' PMT corresponding to each alpha source position in these two runs is given Table~\ref{tab:repeat_BarBRun1} and are consistent within $\sim1\sigma$.  Similarly, Table~\ref{tab:repeat_BarARun2} gives the mean number of photoelectrons measured by the 2'' PMT corresponding to three of the five alpha sources over a span of several hours for bar A, run 1.  The largest drop in the measured number of photoelectrons is 4.3\% and occurs over a span of 981 minutes, which is twice as long as the time required to acquire data corresponding to all five sources.  Conservatively, we apply a 4.5\% systematic error for this run to account for this variation.    

\begin{table}[h]
\centering
\begin{tabular}{ c c c c }
&  \multicolumn{2}{c}{Mean number of photoelectrons} & \\
Alpha position & first measurement & second measurement & time elapsed (days)\\
\hline
1 & 21.24$\pm$0.20 & 21.17$\pm$0.20 & 10.0 \\ 
2 & 20.31$\pm$0.19 & 20.06$\pm$0.18 & 10.4 \\ 
3 & 15.88$\pm$0.15 & 15.52$\pm$0.14 & 10.1 \\ 
4 & 13.91$\pm$0.13 & 13.73$\pm$0.13 & 11.2 \\ 
5 & 12.24$\pm$0.11 & 12.30$\pm$0.11 & 10.8     
\end{tabular}
\caption{Repeat measurements for bar B, run 1 of the mean number of photoelectrons measured by the 2'' PMT corresponding to each alpha position, numbered sequentially starting with the one closest to the photodetector.}
\label{tab:repeat_BarBRun1}
\end{table}

\begin{table}[h]
\centering
\begin{tabular}{ c c c c }
&  \multicolumn{2}{c}{Mean number of photoelectrons} & \\
Alpha positon & first measurement & second measurement & time elapsed (mins)\\
\hline
2 & 17.25$\pm$0.07 & 16.53$\pm$0.07 & 981 \\
3 & 13.72$\pm$0.06 & 13.41$\pm$0.06 & 708 \\
4 & 13.19$\pm$0.06 & 13.07$\pm$0.06 & 209 \\
\end{tabular}
\caption{Repeat measurements for bar A, run 1 of the mean number of photoelectrons measured by the 2'' PMT corresponding to each alpha position, numbered sequentially as above.} 
\label{tab:repeat_BarARun2}
\end{table}

\section{Connecting Attenuation Length Measurements in Air and Liquid Argon}
\label{sec:model}

We use a ray-tracing calculation that includes the effects of partial and total internal reflection, bulk attenuation, and surface losses to model the propagation of light in a TPB-coated, wavelength-shifting acrylic light guide.  We take an approach similar to Ref.~\cite{Jones:2013ks} and assume that the attenuation behavior of light guide bars comes primarily from surface losses, which dominate over bulk attenuation effects.  Surface losses are assumed to be the same in air and in liquid argon, but disparate attenuation behaviors can arise due to the different indices of refraction of air and liquid argon.  We also model a variation in surface coating thickness as a light yield that varies linearly over the length of the bar.  However, we hypothesize that this effect only manifests itself in our air attenuation measurements that use a 286 nm LED because the penetration depth of 128 nm liquid argon scintillation light is too short to be sensitive to coating thickness variations.

\subsection{Fits to Air and Liquid Argon Data}

As illustrated in Figure~\ref{fig:301-A-1_direction}, the attenuation behavior of our light guide bars when measured in air with a 286 nm LED depends on its orientation.  By simultaneously fitting our model to the ``forward'' and ``backward'' measurements in air, we can pin down the effect of a linearly varying light yield and extract the fractional surface loss per bounce $\alpha$.  We can then use this same value of $\alpha$ to predict the attenuation behavior in the liquid argon data, allowing only the overall normalization of the curve to float in the fit.

\begin{figure}[h]
\centering
\includegraphics[width=0.49\textwidth]{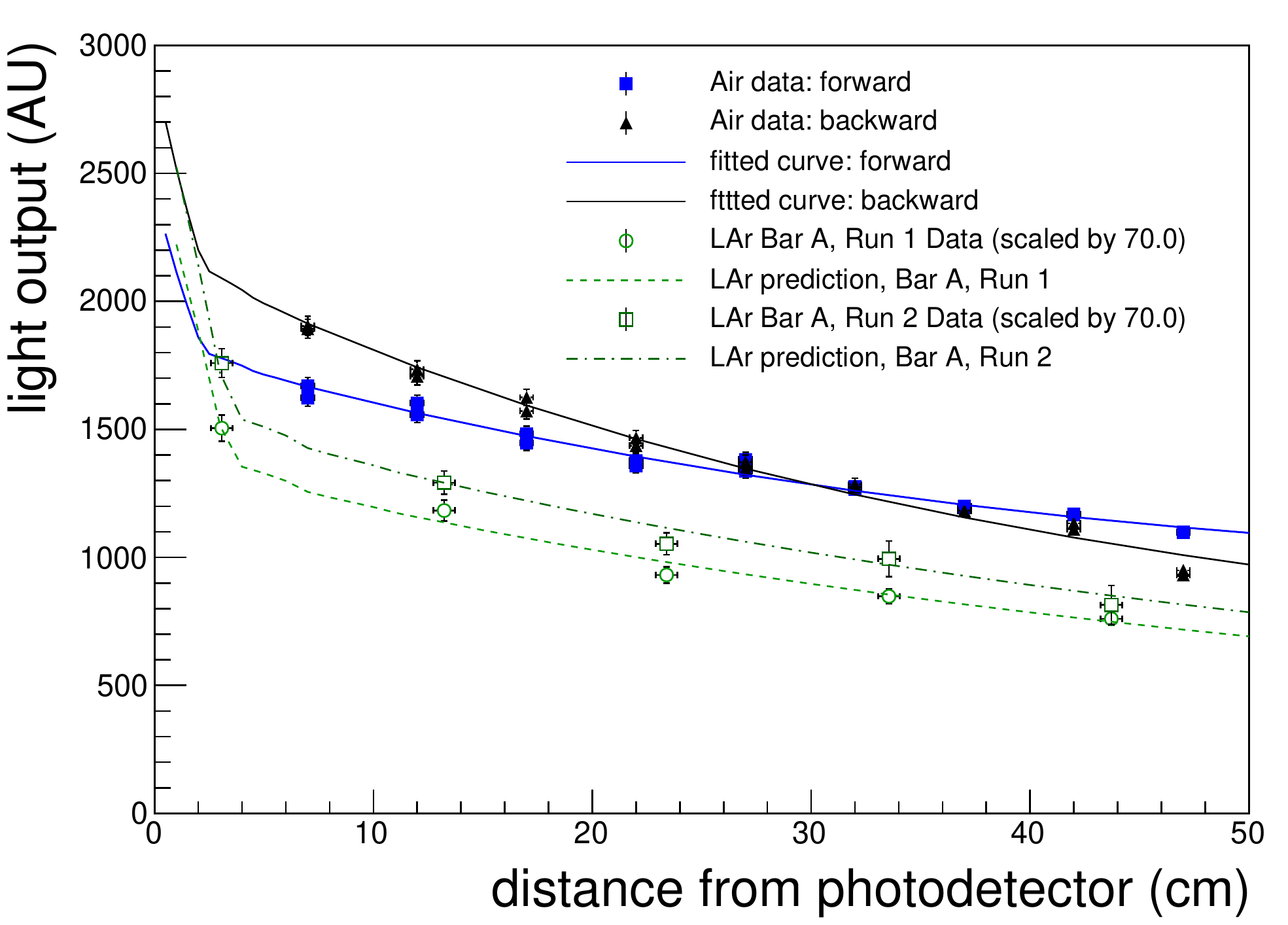} \includegraphics[width=0.49\textwidth]{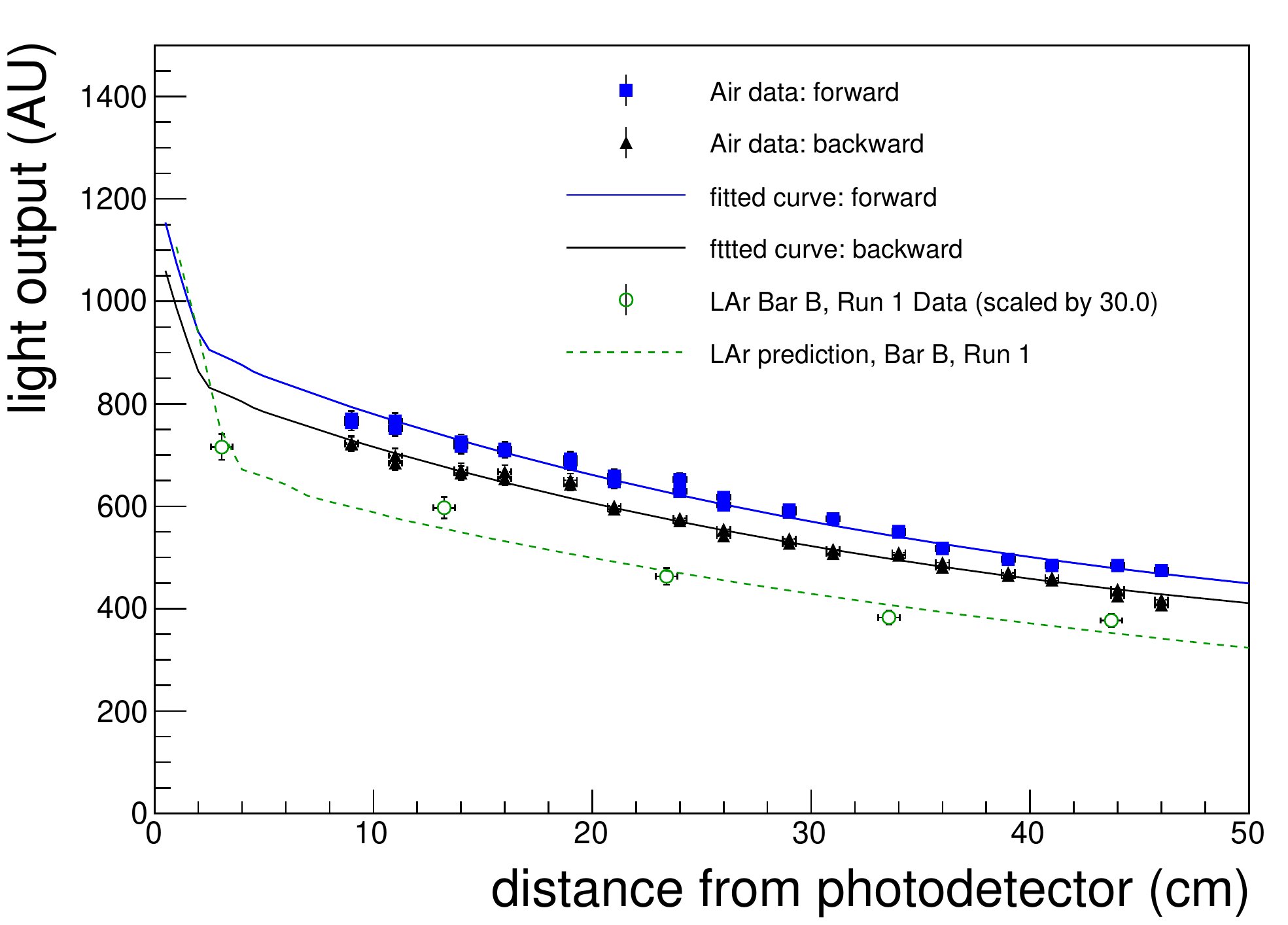}
\caption{\textit{Left:} Model fit to the attenuation data measured in air and liquid argon for bar A.  \textit{Right:} Model fit to the attenuation data measured in air and liquid argon for bar B.}
\label{fig:air_lar_benchmarking}
\end{figure}

Figure~\ref{fig:air_lar_benchmarking} shows the results of the fits of the model to the air and liquid argon attenuation data for bars A and B, respectively.  An exponential function is a worse fit to the data than the model as indicated by comparing the $\chi^2/\textrm{d.o.f.}$ values given previously in Table~\ref{tab:expo_fits} with those reported in Table~\ref{tab:model_fitresults}.  In addition, Table~\ref{tab:model_fitresults} reports the $\chi^2/\textrm{d.o.f.}$ of model fits to liquid argon data that use the linearly varying light yield parameter extracted from fits to air data, which are also found to give a worse fit.

\begin{table}
\centering
\begin{tabular}{ c c c c c }
&   &         &               \multicolumn{2}{c}{$\chi^2$/d.o.f. } \\
Bar & Run & $\alpha$ & w/o coating effect & w/ coating effect \\
\hline
A & 1 & 0.021  & 1.5/4 (3.9/4)  & 6.29/4 (15.8/4) \\ 
A & 2 & 0.021  & 3.9/4  & 12.9/4 \\ 
B & 1 &  0.023 &  10.5/4 & N/A \\ 
\end{tabular}
\caption{For bar A run 1, the values in parenthesis are when
  additional 4.5\% systematic uncertainty due to the purity is removed.} 
\label{tab:model_fitresults}
\end{table}

We note that though the measured attenuation behavior of bars A and B were very different in air, their behavior in liquid argon was similar.  Without the model, it would be difficult to claim that benchmark measurements in air could be related to the behavior of the bars in liquid argon.  However, by using the forward-backward air data, the model is able to account for the coating gradient (or lack thereof) and give a reasonably accurate description of the attenuation data obtained in liquid argon.

%% file: jinst_dipcoating.bbl
\providecommand{\href}[2]{#2}\begingroup\raggedright\begin{thebibliography}{10}

\bibitem{Bugel:2011fw}
L.~Bugel, J.~M. Conrad, C.~Ignarra, B.~J.~P. Jones, T.~Katori, T.~Smidt, and
  H.~K. Tanaka, {\it {Demonstration of a lightguide detector for liquid argon
  TPCs}},  {\em Nuclear Inst. and Methods in Physics Research, A} {\bf 640}
  (2011), no.~1 69--75.

\bibitem{Baptista2013}
B.~Baptista and S.~Mufson, {\it Comparison of tpb and bis-msb as vuv
  waveshifters in prototype lbne photon detector paddles},  {\em JINST} {\bf 8}
  (2013), no.~12 C12003.

\bibitem{Buchanon2013}
R.~Wasserman and N.~Buchanan, {\it Development of a wavelength-shifting
  fiber-based photon detector for lbne},  {\em JINST} {\bf 8} (2013), no.~10
  C10008.

\bibitem{Szelc:2013ht}
A.~M. Szelc, {\it {The LArIAT light readout system}},  {\em JINST} {\bf 8}
  (Sept., 2013) C09011--C09011.

\bibitem{Adams:2013uaa}
{\bf LAr1ND} Collaboration, C.~Adams and {others}, {\it {LAr1-ND: Testing
  Neutrino Anomalies with Multiple LArTPC Detectors at Fermilab}},  {\em
  arXiv.org} {\bf physics.ins-det} (2013)
  [\href{http://arxiv.org/abs/1309.7987}{{\tt arXiv:1309.7987}}].

\bibitem{LBNEdesign}
{\bf LBNE} Collaboration, {\it Volume 4: The liquid argon detector at the far
  site},  tech. rep., LBNE doc 4892,
  http://lbne.fnal.gov/reviews/CD1-CDR.shtml, 2014.

\bibitem{Gehman:2011xm}
V.~M. Gehman, S.~R. Seibert, K.~Rielage, A.~Hime, Y.~Sun, D.~M. Mei,
  J.~Maassen, and D.~Moore, {\it {Nuclear Instruments and Methods in Physics
  Research A}},  {\em Nuclear Inst. and Methods in Physics Research, A} {\bf
  A654} (Oct., 2011) 116--121.

\bibitem{Baptista:2012bf}
B.~Baptista, L.~Bugel, C.~Chiu, J.~M. Conrad, C.~M. Ignarra, and {others}, {\it
  {Benchmarking TPB-coated Light Guides for Liquid Argon TPC Light Detection
  Systems}},  {\em arXiv.org} {\bf physics.ins-det} (2012)
  [\href{http://arxiv.org/abs/1210.3793}{{\tt arXiv:1210.3793}}].

\bibitem{UTRANspec}
L.~International, {\it lucitelux utran\textregistered technical bulletin},
  tech. rep.,
  "http://lucitelux.com/wp-content/uploads/2014/02/LuciteLux\_Utran\_TechnicalBulletin.pdf",
  2014.

\bibitem{Andrews:2009zza}
R.~Andrews, W.~Jaskierny, H.~Jostlein, C.~Kendziora, S.~Pordes, and {others},
  {\it {A system to test the effects of materials on the electron drift
  lifetime in liquid argon and observations on the effect of water}},  {\em
  Nuclear Inst. and Methods in Physics Research, A} {\bf A608} (2009), no.~2
  251--258.

\bibitem{Jones:2013bca}
B.~Jones, C.~Chiu, J.~Conrad, C.~Ignarra, T.~Katori, et~al., {\it {A
  Measurement of the Absorption of Liquid Argon Scintillation Light by
  Dissolved Nitrogen at the Part-Per-Million Level}},  {\em JINST} {\bf 8}
  (2013) P07011, [\href{http://arxiv.org/abs/1306.4605}{{\tt
  arXiv:1306.4605}}].

\bibitem{Jones:2013mfa}
B.~Jones, T.~Alexander, H.~Back, G.~Collin, J.~Conrad, et~al., {\it {The
  Effects of Dissolved Methane upon Liquid Argon Scintillation Light}},  {\em
  JINST} {\bf 8} (2013) P12015, [\href{http://arxiv.org/abs/1308.3658}{{\tt
  arXiv:1308.3658}}].

\bibitem{Conrad:2015xta}
J.~Conrad, B.~Jones, Z.~Moss, T.~Strauss, and M.~Toups, {\it {The
  Photomultiplier Tube Calibration System of the MicroBooNE Experiment}},  {\em
  arXiv.org} (2015) [\href{http://arxiv.org/abs/1502.0415}{{\tt
  arXiv:1502.0415}}].

\bibitem{WArP_nitrogen}
{\bf WArP Collaboration} Collaboration, R.~Acciarri et~al., {\it {Effects of
  Nitrogen contamination in liquid Argon}},  {\em JINST} {\bf 5} (2010) P06003,
  [\href{http://arxiv.org/abs/0804.1217}{{\tt arXiv:0804.1217}}].

\bibitem{WArP_oxygen}
{\bf WArP Collaboration} Collaboration, R.~Acciarri et~al., {\it {Oxygen
  contamination in liquid Argon: Combined effects on ionization electron charge
  and scintillation light}},  {\em JINST} {\bf 5} (2010) P05003,
  [\href{http://arxiv.org/abs/0804.1222}{{\tt arXiv:0804.1222}}].

\bibitem{Jones:2013ks}
B.~J.~P. Jones, {\it {A simulation of the optical attenuation of TPB coated
  light-guide detectors}},  {\em JINST} {\bf 8} (Oct., 2013) C10015.

\end{thebibliography}\endgroup
